 %%%%%%%%%%%%%%%%%%     PLAIN TEX FILE       %%%%%%%%%%%%%%%%%%
 %%%%%%%%%%%%%%%%%%  %%%%%%%%%%%%%%%%%%  %%%%%%%%%%%%%%%%%%  %%%%%%%%%%%%%%%%%%
 %%%%%%%%%%%%%%%%%%  %%%%%%%%%%%%%%%%%%  %%%%%%%%%%%%%%%%%%  %%%%%%%%%%%%%%%%%%
 %%%%%%%%%%%%%%%%%%  %%%%%%%%%%%%%%%%%%  %%%%%%%%%%%%%%%%%%  %%%%%%%%%%%%%%%%%%

 %%%%%%%%%%%%%%%%%%  tex macros for preprints, cm version %%%%%%%%%%%%%%
%                     (P. Ginsparg, last updated 9/91)
%                if confused, type `b' in response to query
%
%---------------------------------------------------------------------%
%% site dependent options:
%% \unredoffs and \redoffs define horizontal and vertical offsets
%% respectively for unreduced and reduced modes. \speclscape defines
%% the \special{} call that sets printer to landscape (sideways) mode.
%% from standard set below, leave uncommented as appropriate or redefine
%
%%% next 400dpi
%\def\unredoffs{} \def\redoffs{\voffset=-.31truein\hoffset=-.48truein}
%\def\speclscape{\special{landscape}}
%
%%% apple lw
\def\unredoffs{} \def\redoffs{\voffset=-.31truein\hoffset=-.59truein}
\def\speclscape{\special{ps: landscape}}
%
%%% qms lasergrafix:
%\def\unredoffs{} \def\redoffs{\voffset=-.4truein\hoffset=.125truein}
%\def\speclscape{\special{qms: landscape}}
%
%%% saclay A4 paper:
%\def\unredoffs{\hoffset-.14truein\voffset-.2truein}
%\def\redoffs{\voffset=-.55truein\hoffset=-.1truein} \def\speclscape{}
%
%---------------------------------------------------------------------%
%
\newbox\leftpage \newdimen\fullhsize \newdimen\hstitle \newdimen\hsbody
\tolerance=1000\hfuzz=2pt
\catcode`\@=11 % This allows us to modify PLAIN macros.
\def\bigans{b }
%\message{ big or little (b/l)? }\read-1 to\answ
\def\answ{b }
\ifx\answ\bigans\message{(This will come out unreduced.}
\magnification=1200\unredoffs\baselineskip=16pt plus 2pt minus 1pt
\hsbody=\hsize \hstitle=\hsize %take default values for unreduced format
\else\message{(This will be reduced.} \let\l@r=L
\magnification=1000\baselineskip=16pt plus 2pt minus 1pt \vsize=7truein
\redoffs \hstitle=8truein\hsbody=4.75truein\fullhsize=10truein\hsize=\hsbody
\output={\ifnum\pageno=0 %%% This is the HUTP version
  \shipout\vbox{\speclscape{\hsize\fullhsize\makeheadline}
    \hbox to \fullhsize{\hfill\pagebody\hfill}}\advancepageno
  \else
  \almostshipout{\leftline{\vbox{\pagebody\makefootline}}}\advancepageno
  \fi}
\def\almostshipout#1{\if L\l@r \count1=1 \message{[\the\count0.\the\count1]}
      \global\setbox\leftpage=#1 \global\let\l@r=R
 \else \count1=2
  \shipout\vbox{\speclscape{\hsize\fullhsize\makeheadline}
      \hbox to\fullhsize{\box\leftpage\hfil#1}}  \global\let\l@r=L\fi}
\fi
%---------------------------------------------------------------------
%
\newcount\yearltd\yearltd=\year\advance\yearltd by -1900

\def\Title#1#2{\nopagenumbers\abstractfont\hsize=\hstitle\rightline{#1}%
\vskip 1in\centerline{\titlefont #2}\abstractfont\vskip .5in\pageno=0}
\def\Date#1{\vfill\leftline{#1}\tenpoint\supereject\global\hsize=\hsbody%
\footline={\hss\tenrm\folio\hss}}%      restores pagenumbers
%
%       use following instead of \Date on the preliminary draft,
%       puts date/time on each page in big mode, writes labels in margins

\def\draftmode{\message{ DRAFTMODE }\def\draftdate{{\rm preliminary draft:
\number\month/\number\day/\number\yearltd\ \ \hourmin}}%
\headline={\hfil\draftdate}\writelabels\baselineskip=20pt plus 2pt minus 2pt
 {\count255=\time\divide\count255 by 60 \xdef\hourmin{\number\count255}
  \multiply\count255 by-60\advance\count255 by\time
  \xdef\hourmin{\hourmin:\ifnum\count255<10 0\fi\the\count255}}}
%       use \nolabels to get rid of eqn, ref, and fig labels in draft mode
\def\nolabels{\def\wrlabeL##1{}\def\eqlabeL##1{}\def\reflabeL##1{}}
\def\writelabels{\def\wrlabeL##1{\leavevmode\vadjust{\rlap{\smash%
{\line{{\escapechar=` \hfill\rlap{\sevenrm\hskip.03in\string##1}}}}}}}%
\def\eqlabeL##1{{\escapechar-1\rlap{\sevenrm\hskip.05in\string##1}}}%
\def\reflabeL##1{\noexpand\llap{\noexpand\sevenrm\string\string\string##1}}}
\nolabels
%
% tagged sec numbers
\global\newcount\secno \global\secno=0
\global\newcount\meqno \global\meqno=1
\def\newsec#1{\global\advance\secno by1\message{(\the\secno. #1)}
%\ifx\answ\bigans \vfill\eject \else \bigbreak\bigskip \fi  %if desired
\global\subsecno=0\eqnres@t\noindent{\bf\the\secno. #1}
\writetoca{{\secsym} {#1}}\par\nobreak\medskip\nobreak}
\def\eqnres@t{\xdef\secsym{\the\secno.}\global\meqno=1\bigbreak\bigskip}
\def\sequentialequations{\def\eqnres@t{\bigbreak}}\xdef\secsym{}
\global\newcount\subsecno \global\subsecno=0
\def\subsec#1{\global\advance\subsecno by1\message{(\secsym\the\subsecno.
#1)}
\ifnum\lastpenalty>9000\else\bigbreak\fi
\noindent{\it\secsym\the\subsecno. #1}\writetoca{\string\quad
{\secsym\the\subsecno.} {#1}}\par\nobreak\medskip\nobreak}
\def\appendix#1#2{\global\meqno=1\global\subsecno=0\xdef\secsym{\hbox{#1.}}
\bigbreak\bigskip\noindent{\bf Appendix #1. #2}\message{(#1. #2)}
\writetoca{Appendix {#1.} {#2}}\par\nobreak\medskip\nobreak}
%
%       \eqn\label{a+b=c}       gives displayed equation, numbered
%                               consecutively within sections.
%     \eqnn and \eqna define labels in advance (of eqalign?)
%
\def\eqnn#1{\xdef #1{(\secsym\the\meqno)}\writedef{#1\leftbracket#1}%
\global\advance\meqno by1\wrlabeL#1}
\def\eqna#1{\xdef #1##1{\hbox{$(\secsym\the\meqno##1)$}}
\writedef{#1\numbersign1\leftbracket#1{\numbersign1}}%
\global\advance\meqno by1\wrlabeL{#1$\{\}$}}
\def\eqn#1#2{\xdef #1{(\secsym\the\meqno)}\writedef{#1\leftbracket#1}%
\global\advance\meqno by1$$#2\eqno#1\eqlabeL#1$$}
%
%                            footnotes
\newskip\footskip\footskip14pt plus 1pt minus 1pt %sets footnote baselineskip
\def\footnotefont{\ninepoint}\def\f@t#1{\footnotefont #1\@foot}
\def\f@@t{\baselineskip\footskip\bgroup\footnotefont\aftergroup\@foot\let\next}
\setbox\strutbox=\hbox{\vrule height9.5pt depth4.5pt width0pt}
\global\newcount\ftno \global\ftno=0
\def\foot{\global\advance\ftno by1\footnote{$^{\the\ftno}$}}
%
%say \footend to put footnotes at end
%will cause problems if \ref used inside \foot, instead use \nref before
\newwrite\ftfile
\def\footend{\def\foot{\global\advance\ftno by1\chardef\wfile=\ftfile
$^{\the\ftno}$\ifnum\ftno=1\immediate\openout\ftfile=foots.tmp\fi%
\immediate\write\ftfile{\noexpand\smallskip%
\noexpand\item{f\the\ftno:\ }\pctsign}\findarg}%
\def\footatend{\vfill\eject\immediate\closeout\ftfile{\parindent=20pt
\centerline{\bf Footnotes}\nobreak\bigskip\input foots.tmp }}}
\def\footatend{}
%
%     \ref\label{text}
% generates a number, assigns it to \label, generates an entry.
% To list the refs on a separate page,  \listrefs
%
\global\newcount\refno \global\refno=1
\newwrite\rfile
\def\ref{[\the\refno]\nref}
\def\nref#1{\xdef#1{[\the\refno]}\writedef{#1\leftbracket#1}%
\ifnum\refno=1\immediate\openout\rfile=refs.tmp\fi
\global\advance\refno by1\chardef\wfile=\rfile\immediate
\write\rfile{\noexpand\item{#1\ }\reflabeL{#1\hskip.31in}\pctsign}\findarg}
%        horrible hack to sidestep tex \write limitation
\def\findarg#1#{\begingroup\obeylines\newlinechar=`\^^M\pass@rg}
{\obeylines\gdef\pass@rg#1{\writ@line\relax #1^^M\hbox{}^^M}%
\gdef\writ@line#1^^M{\expandafter\toks0\expandafter{\striprel@x #1}%
\edef\next{\the\toks0}\ifx\next\em@rk\let\next=\endgroup\else\ifx\next\empty%
\else\immediate\write\wfile{\the\toks0}\fi\let\next=\writ@line\fi\next\relax}}
\def\striprel@x#1{} \def\em@rk{\hbox{}}
\def\lref{\begingroup\obeylines\lr@f}
\def\lr@f#1#2{\gdef#1{\ref#1{#2}}\endgroup\unskip}
\def\semi{;\hfil\break}
\def\addref#1{\immediate\write\rfile{\noexpand\item{}#1}} %now unnecessary
\def\startrefs#1{\immediate\openout\rfile=refs.tmp\refno=#1}
\def\xref{\expandafter\xr@f}\def\xr@f[#1]{#1}
\def\refs#1{\count255=1[\r@fs #1{\hbox{}}]}
\def\r@fs#1{\ifx\und@fined#1\message{reflabel \string#1 is undefined.}%
\nref#1{need to supply reference \string#1.}\fi%
\vphantom{\hphantom{#1}}\edef\next{#1}\ifx\next\em@rk\def\next{}%
\else\ifx\next#1\ifodd\count255\relax\xref#1\count255=0\fi%
\else#1\count255=1\fi\let\next=\r@fs\fi\next}
%

%
% this is ugly, but moore insists
\newwrite\ffile\global\newcount\figno \global\figno=1
\def\fig{fig.~\the\figno\nfig}
\def\nfig#1{\xdef#1{fig.~\the\figno}%
\writedef{#1\leftbracket fig.\noexpand~\the\figno}%
\ifnum\figno=1\immediate\openout\ffile=figs.tmp\fi\chardef\wfile=\ffile%
\immediate\write\ffile{\noexpand\medskip\noexpand\item{Fig.\ \the\figno. }
\reflabeL{#1\hskip.55in}\pctsign}\global\advance\figno by1\findarg}
\def\xfig{\expandafter\xf@g}\def\xf@g fig.\penalty\@M\ {}
\def\figs#1{figs.~\f@gs #1{\hbox{}}}
\def\f@gs#1{\edef\next{#1}\ifx\next\em@rk\def\next{}\else
\ifx\next#1\xfig #1\else#1\fi\let\next=\f@gs\fi\next}
\newwrite\lfile
{\escapechar-1\xdef\pctsign{\string\%}\xdef\leftbracket{\string\{}
\xdef\rightbracket{\string\}}\xdef\numbersign{\string\#}}

\def\writestop{\def\writestoppt{\immediate\write\lfile{\string\pageno%
\the\pageno\string\startrefs\leftbracket\the\refno\rightbracket%
\string\def\string\secsym\leftbracket\secsym\rightbracket%
\string\secno\the\secno\string\meqno\the\meqno}\immediate\closeout\lfile}}
\def\writestoppt{}\def\writedef#1{}
\def\seclab#1{\xdef #1{\the\secno}\writedef{#1\leftbracket#1}\wrlabeL{#1=#1}}
\def\subseclab#1{\xdef #1{\secsym\the\subsecno}%
\writedef{#1\leftbracket#1}\wrlabeL{#1=#1}}
\newwrite\tfile \def\writetoca#1{}
\def\leaderfill{\leaders\hbox to 1em{\hss.\hss}\hfill}
%        use this to write file with table of contents
\def\writetoc{\immediate\openout\tfile=toc.tmp
   \def\writetoca##1{{\edef\next{\write\tfile{\noindent ##1
   \string\leaderfill {\noexpand\number\pageno} \par}}\next}}}
%       and this lists table of contents on second pass

%
\catcode`\@=12 % at signs are no longer letters
%
%        Unpleasantness in calling in abstract and title fonts
\edef\tfontsize{\ifx\answ\bigans scaled\magstep3\else scaled\magstep4\fi}
\font\titlerm=cmr10 \tfontsize \font\titlerms=cmr7 \tfontsize
\font\titlermss=cmr5 \tfontsize \font\titlei=cmmi10 \tfontsize
\font\titleis=cmmi7 \tfontsize \font\titleiss=cmmi5 \tfontsize
\font\titlesy=cmsy10 \tfontsize \font\titlesys=cmsy7 \tfontsize
\font\titlesyss=cmsy5 \tfontsize \font\titleit=cmti10 \tfontsize
\skewchar\titlei='177 \skewchar\titleis='177 \skewchar\titleiss='177
\skewchar\titlesy='60 \skewchar\titlesys='60 \skewchar\titlesyss='60
\def\titlefont{\def\rm{\fam0\titlerm}% switch to title font
\textfont0=\titlerm \scriptfont0=\titlerms \scriptscriptfont0=\titlermss
\textfont1=\titlei \scriptfont1=\titleis \scriptscriptfont1=\titleiss
\textfont2=\titlesy \scriptfont2=\titlesys \scriptscriptfont2=\titlesyss
\textfont\itfam=\titleit \def\it{\fam\itfam\titleit}\rm}
 \ifx\answ\bigans\else scaled\magstep1\fi
\ifx\answ\bigans\def\abstractfont{\tenpoint}\else
\font\abssl=cmsl10 scaled \magstep1
\font\absrm=cmr10 scaled\magstep1 \font\absrms=cmr7 scaled\magstep1
\font\absrmss=cmr5 scaled\magstep1 \font\absi=cmmi10 scaled\magstep1
\font\absis=cmmi7 scaled\magstep1 \font\absiss=cmmi5 scaled\magstep1
\font\abssy=cmsy10 scaled\magstep1 \font\abssys=cmsy7 scaled\magstep1
\font\abssyss=cmsy5 scaled\magstep1 \font\absbf=cmbx10 scaled\magstep1
\skewchar\absi='177 \skewchar\absis='177 \skewchar\absiss='177
\skewchar\abssy='60 \skewchar\abssys='60 \skewchar\abssyss='60
\def\abstractfont{\def\rm{\fam0\absrm}% switch to abstract font
\textfont0=\absrm \scriptfont0=\absrms \scriptscriptfont0=\absrmss
\textfont1=\absi \scriptfont1=\absis \scriptscriptfont1=\absiss
\textfont2=\abssy \scriptfont2=\abssys \scriptscriptfont2=\abssyss
\textfont\itfam=\bigit \def\it{\fam\itfam\bigit}\def\footnotefont{\tenpoint}%
\textfont\slfam=\abssl \def\sl{\fam\slfam\abssl}%
\textfont\bffam=\absbf \def\bf{\fam\bffam\absbf}\rm}\fi
\def\tenpoint{\def\rm{\fam0\tenrm}% switch back to 10-point type
\textfont0=\tenrm \scriptfont0=\sevenrm \scriptscriptfont0=\fiverm
\textfont1=\teni  \scriptfont1=\seveni  \scriptscriptfont1=\fivei
\textfont2=\tensy \scriptfont2=\sevensy \scriptscriptfont2=\fivesy
\textfont\itfam=\tenit
\def\it{\fam\itfam\tenit}\def\footnotefont{\ninepoint}%
\textfont\bffam=\tenbf \def\bf{\fam\bffam\tenbf}\def\sl{\fam\slfam\tensl}\rm}
\font\ninerm=cmr9 \font\sixrm=cmr6 \font\ninei=cmmi9 \font\sixi=cmmi6
\font\ninesy=cmsy9 \font\sixsy=cmsy6 \font\ninebf=cmbx9
\font\nineit=cmti9 \font\ninesl=cmsl9 \skewchar\ninei='177
\skewchar\sixi='177 \skewchar\ninesy='60 \skewchar\sixsy='60
\def\ninepoint{\def\rm{\fam0\ninerm}% switch to footnote font
\textfont0=\ninerm \scriptfont0=\sixrm \scriptscriptfont0=\fiverm
\textfont1=\ninei \scriptfont1=\sixi \scriptscriptfont1=\fivei
\textfont2=\ninesy \scriptfont2=\sixsy \scriptscriptfont2=\fivesy
\textfont\itfam=\ninei \def\it{\fam\itfam\nineit}\def\sl{\fam\slfam\ninesl}%
\textfont\bffam=\ninebf \def\bf{\fam\bffam\ninebf}\rm}
%
%---------------------------------------------------------------------
%

\hyphenation{anom-aly anom-alies coun-ter-term coun-ter-terms}
\def\inv{^{\raise.15ex\hbox{${\scriptscriptstyle -}$}\kern-.05em 1}}

\def\Dsl{\,\raise.15ex\hbox{/}\mkern-13.5mu D} %this one can be subscripted
\def\dsl{\raise.15ex\hbox{/}\kern-.57em\partial}

\font\bigit=cmti10 scaled \magstep1
 %pound sterling
\def\lspace{\ifx\answ\bigans{}\else\qquad\fi}
\def\lbspace{\ifx\answ\bigans{}\else\hskip-.2in\fi} % $$\lbspace...$$
\def\boxeqn#1{\vcenter{\vbox{\hrule\hbox{\vrule\kern3pt\vbox{\kern3pt
           \hbox{${\displaystyle #1}$}\kern3pt}\kern3pt\vrule}\hrule}}}
\def\mbox#1#2{\vcenter{\hrule \hbox{\vrule height#2in
               \kern#1in \vrule} \hrule}}  %e.g. \mbox{.1}{.1}
%       matters of taste
%\def\tilde{\widetilde} \def\bar{\overline} \def\hat{\widehat}
%
% some sample definitions
  %     curly letters

\def\darr#1{\raise1.5ex\hbox{$\leftrightarrow$}\mkern-16.5mu #1}
 %pound sterling

 %puts a small half in a displayed eqn
\def\roughly#1{\raise.3ex\hbox{$#1$\kern-.75em\lower1ex\hbox{$\sim$}}}

\input epsf.tex

%%temporary additional macros
% \input macros.tex
% April 16 -- NN

%%%%%%%%%%%%%%%%%%%%%  Rublenye bukvy   %%%%%%%%%%%%%%%%%%%%%%%%
\def\IB{\relax\hbox{$\inbar\kern-.3em{\rm B}$}}
\def\IC{\relax\hbox{$\inbar\kern-.3em{\rm C}$}}
\def\ID{\relax\hbox{$\inbar\kern-.3em{\rm D}$}}
\def\IE{\relax\hbox{$\inbar\kern-.3em{\rm E}$}}
\def\IF{\relax\hbox{$\inbar\kern-.3em{\rm F}$}}
\def\IG{\relax\hbox{$\inbar\kern-.3em{\rm G}$}}
\def\IGa{\relax\hbox{${\rm I}\kern-.18em\Gamma$}}
\def\IH{\relax{\rm I\kern-.18em H}}
\def\IK{\relax{\rm I\kern-.18em K}}
\def\II{\relax{\rm I\kern-.18em I}}
\def\IL{\relax{\rm I\kern-.18em L}}
\def\IP{\relax{\rm I\kern-.18em P}}
\def\IR{\relax{\rm I\kern-.18em R}}
\def\IZ{\relax\ifmmode\mathchoice {\hbox{\cmss Z\kern-.4em Z}}{\hbox{\cmss
Z\kern-.4em Z}} {\lower.9pt\hbox{\cmsss Z\kern-.4em Z}}
{\lower1.2pt\hbox{\cmsss Z\kern-.4em Z}}\else{\cmss Z\kern-.4em Z}\fi}

\def\IB{\relax{\rm I\kern-.18em B}}
\def\IC{{\relax\hbox{$\inbar\kern-.3em{\rm C}$}}}
\def\ID{\relax{\rm I\kern-.18em D}}
\def\IE{\relax{\rm I\kern-.18em E}}
\def\IF{\relax{\rm I\kern-.18em F}}

%%%%%%%%%%%%%%%%%%%% Calligraphic letters  %%%%%%%%%%%%%%%%%%%%%%%

\def\CW {{\cal W}}

%%%%%%%%%%%%%%%%%%%%%%%%%% Derivatives  %%%%%%%%%%%%%%%%%%%%%%%%
\def\p{\partial}

%%Beltrami

%%%%%%%%%%%%%%%%%%%% letters with bar %%%%%%%%%%%%%%%%%%%%%%%%%%

%%%%%%%%%%%%%%%%%%%%%%%%%%% Math symbols %%%%%%%%%%%%%%%%%%%%%%%

%%%%%%%%%%%%%%%%%%%%% Short Cuts %%%%%%%%%%%%%%%%%%%%%%%

\def\demi{{1\over 2}}

\def\c{\cdot}

%%%%%%%%%%%%%%%%%% Greek %%%%%%%%%%%%%%%%%%%%%%

\def\f{\phi}

\def\a{\alpha}
\def\b{\beta}
  
\def\d{\delta}  
\def\m{\mu}
\def\n{\nu}
\def\r{\rho}
\def\l{\lambda} \def\L{\Lambda}

%%%%%%%%%%%%%%%%%% Big ( )  %%%%%%%%%%%%%%%%%%%%%%
\def\|{\Big|}
\def\({\Big(}   \def\){\Big)}
\def\[{\Big[}   \def\]{\Big]}

%%%%%%%%%%%%%%%%%% Text %%%%%%%%%%%%%%%%%%%%%%

%%%%%%%%%%%%% References %%%%%%%%%%%%%%%%%%%%

\def\paper#1#2#3#4{#1, {\sl #2}, #3 {\tt #4}}
% refs with #1=authors, #2=title, #3=publ.ref, #4=hep no :
%\lref\NAME{\paper
%{Authors}{Title(in \it)}{\PLB{No.}{Year}{page},}
%{\hh 0008248 (in\tt)}.}

%\def\hh#1{hep-th/{\it #1}}
\def\hh{hep-th/}

% journal~{\bf no.} (year) page

\def\PLB#1#2#3{Phys. Lett.~{\bf B#1} (#2) #3}
\def\NPB#1#2#3{Nucl. Phys.~{\bf B#1} (#2) #3}
\def\PRL#1#2#3{Phys. Rev. Lett.~{\bf #1} (#2) #3}
\def\CMP#1#2#3{Comm. Math. Phys.~{\bf #1} (#2) #3}
\def\PRD#1#2#3{Phys. Rev.~{\bf D#1} (#2) #3}
\def\MPL#1#2#3{Mod. Phys. Lett.~{\bf #1} (#2) #3}
\def\IJMP#1#2#3{Int. Jour. Mod. Phys.~{\bf #1} (#2) #3}

%%%%%%%%%%%%%%%%%%% Something to deal with sub-sub-sections
%%%%%%%%%%%%%%%%%%%%%%%%%%%%%%%%%%%%%%%%%%%%%%%

\def\unlockat{\catcode`\@=11}
\def\lockat{\catcode`\@=12}

\unlockat

% Something to deal with sub-sub-sections

\def\newsec#1{\global\advance\secno by1\message{(\the\secno. #1)}
\global\subsecno=0\global\subsubsecno=0\eqnres@t\noindent {\bf\the\secno. #1}
\writetoca{{\secsym} {#1}}\par\nobreak\medskip\nobreak}
\global\newcount\subsecno \global\subsecno=0
\def\subsec#1{\global\advance\subsecno by1\message{(\secsym\the\subsecno.
#1)}
\ifnum\lastpenalty>9000\else\bigbreak\fi\global\subsubsecno=0
\noindent{\it\secsym\the\subsecno. #1}
\writetoca{\string\quad {\secsym\the\subsecno.} {#1}}
\par\nobreak\medskip\nobreak}
\global\newcount\subsubsecno \global\subsubsecno=0
\def\subsubsec#1{\global\advance\subsubsecno by1
\message{(\secsym\the\subsecno.\the\subsubsecno. #1)}
\ifnum\lastpenalty>9000\else\bigbreak\fi
\noindent\quad{\secsym\the\subsecno.\the\subsubsecno.}{#1}
\writetoca{\string\qquad{\secsym\the\subsecno.\the\subsubsecno.}{#1}}
\par\nobreak\medskip\nobreak}

\def\subsubseclab#1{\DefWarn#1\xdef #1{\noexpand\hyperref{}{subsubsection}%
{\secsym\the\subsecno.\the\subsubsecno}%
{\secsym\the\subsecno.\the\subsubsecno}}%
\writedef{#1\leftbracket#1}\wrlabeL{#1=#1}}% Macros for boxes
\lockat

%why???\font\manual=manfnt
\def\dbend{\lower3.5pt\hbox{\manual\char127}}

%%%%%%%%%%%%%%%%%%% Macros for boxes %%%%%%%%%%%%%%%%%%

\def\boxit#1{\vbox{\hrule\hbox{\vrule\kern8pt
\vbox{\hbox{\kern8pt}\hbox{\vbox{#1}}\hbox{\kern8pt}}
\kern8pt\vrule}\hrule}}

\def\mathboxit#1{\vbox{\hrule\hbox{\vrule\kern8pt\vbox{\kern8pt
\hbox{$\displaystyle #1$}\kern8pt}\kern8pt\vrule}\hrule}}

%%%%%%%%%%%%%%%%%%%% ANOTHER SET OF MACROS %%%%%%%%%%%%%%%%%%

\def\inbar{\,\vrule height1.5ex width.4pt depth0pt}

\font\cmss=cmss10 \font\cmsss=cmss10 at 7pt

%REFERENCES
%%%%%%%%%%%%%%%%%%%%%%%%%%%%%%%%%%%%%%%%%%%%%%%%%%%

\lref\simons{ J. Cheeger and J. Simons, {\it Differential Characters and
Geometric Invariants},  Stony Brook Preprint, (1973), unpublished.}

\lref\cargese{ L.~Baulieu, {\it Algebraic quantization of gauge theories},
Perspectives in fields and particles, Plenum Press, eds. Basdevant-Levy,
Cargese Lectures 1983}

\lref\antifields{ L. Baulieu, M. Bellon, S. Ouvry, C.Wallet, Phys.Letters
B252 (1990) 387; M.  Bocchichio, Phys. Lett. B187 (1987) 322;  Phys. Lett. B
192 (1987) 31; R.  Thorn    Nucl. Phys.   B257 (1987) 61. }

\lref\thompson{ George Thompson,  Annals Phys. 205 (1991) 130; J.M.F.
Labastida, M. Pernici, Phys. Lett. 212B  (1988) 56; D. Birmingham, M.Blau,
M. Rakowski and G.Thompson, Phys. Rept. 209 (1991) 129.}

\lref\tonin{ Tonin}

\lref\wittensix{ E.  Witten, {\it New  Gauge  Theories In Six Dimensions},
\hh{9710065}. }

\lref\orlando{ O. Alvarez, L. A. Ferreira and J. Sanchez Guillen, {\it  A New
Approach to Integrable Theories in any Dimension}, hep-th/9710147.}

\lref\wittentopo{ E.  Witten,  {\it  Topological Quantum Field Theory},
\hh9403195, Commun.  Math. Phys.  {117} (1988)353.  }

\lref\wittentwist{ E.  Witten, {\it Supersymmetric Yang--Mills theory on a
four-manifold}, J.  Math.  Phys.  {35} (1994) 5101.}

\lref\west{ L.~Baulieu, P.~West, {\it Six Dimensional TQFTs and  Self-dual
Two-Forms,} Phys.Lett. B {\bf 436 } (1998) 97, /hep-th/9805200}

\lref\bv{ I.A. Batalin and V.A. Vilkowisky,    Phys. Rev.   D28  (1983)
2567\semi M. Henneaux,  Phys. Rep.  126   (1985) 1\semi M. Henneaux and C.
Teitelboim, {\it Quantization of Gauge Systems}
  Princeton University Press,  Princeton (1992).}

\lref\kyoto{ L. Baulieu, E. Bergschoeff and E. Sezgin, Nucl. Phys.
B307(1988)348\semi L. Baulieu,   {\it Field Antifield Duality, p-Form Gauge
Fields
   and Topological Quantum Field Theories}, hep-th/9512026,
   Nucl. Phys. B478 (1996) 431.  }

\lref\sourlas{ G. Parisi and N. Sourlas, {\it Random Magnetic Fields,
Supersymmetry and Negative Dimensions}, Phys. Rev. Lett.  43 (1979) 744;
Nucl.  Phys.  B206 (1982) 321.  }

\lref\SalamSezgin{ A.  Salam  and  E.  Sezgin, {\it Supergravities in
diverse dimensions}, vol.  1, p. 119\semi P.  Howe, G.  Sierra and P.
Townsend, Nucl Phys B221 (1983) 331.}

\lref\nekrasov{ A. Losev, G. Moore, N. Nekrasov, S. Shatashvili, {\it
Four-Dimensional Avatars of Two-Dimensional RCFT},  hep-th/9509151, Nucl.
Phys.  Proc.  Suppl.   46 (1996) 130\semi L.  Baulieu, A.  Losev,
N.~Nekrasov  {\it Chern-Simons and Twisted Supersymmetry in Higher
Dimensions},  hep-th/9707174, to appear in Nucl.  Phys.  B.  }

\lref\WitDonagi{R.~ Donagi, E.~ Witten, ``Supersymmetric Yang--Mills Theory
and Integrable Systems'', hep-th/9510101, Nucl. Phys.{\bf B}460 (1996)
299-334}
\lref\Witfeb{E.~ Witten, ``Supersymmetric Yang--Mills Theory On A
Four-Manifold,''  hep-th/9403195; J. Math. Phys. {\bf 35} (1994) 5101.}
\lref\Witgrav{E.~ Witten, ``Topological Gravity'', Phys.Lett.206B:601, 1988}
\lref\witaffl{I. ~ Affleck, J.A.~ Harvey and E.~ Witten,
        ``Instantons and (Super)Symmetry Breaking
        in $2+1$ Dimensions'', Nucl. Phys. {\bf B}206 (1982) 413}
\lref\wittabl{E.~ Witten,  ``On $S$-Duality in Abelian Gauge Theory,''
hep-th/9505186; Selecta Mathematica {\bf 1} (1995) 383}
\lref\wittgr{E.~ Witten, ``The Verlinde Algebra And The Cohomology Of The
Grassmannian'',  hep-th/9312104}
\lref\wittenwzw{E. Witten, ``Non Abelian bosonization in two dimensions,''
Commun. Math. Phys. {\bf 92} (1984)455 }
\lref\witgrsm{E. Witten, ``Quantum field theory, grassmannians and algebraic
curves,'' Commun.Math.Phys.113:529,1988}
\lref\wittjones{E. Witten, ``Quantum field theory and the Jones
polynomial,'' Commun.  Math. Phys., 121 (1989) 351. }
\lref\witttft{E.~ Witten, ``Topological Quantum Field Theory", Commun. Math.
Phys. {\bf 117} (1988) 353.}
\lref\wittmon{E.~ Witten, ``Monopoles and Four-Manifolds'', hep-th/9411102}
\lref\Witdgt{ E.~ Witten, ``On Quantum gauge theories in two dimensions,''
Commun. Math. Phys. {\bf  141}  (1991) 153}
\lref\witrevis{E.~ Witten,
 ``Two-dimensional gauge theories revisited'', hep-th/9204083; J. Geom.
Phys. 9 (1992) 303-368}
\lref\Witgenus{E.~ Witten, ``Elliptic Genera and Quantum Field Theory'',
Comm. Math. Phys. 109(1987) 525. }
\lref\OldZT{E. Witten, ``New Issues in Manifolds of SU(3) Holonomy,'' {\it
Nucl. Phys.} {\bf B268} (1986) 79 \semi J. Distler and B. Greene, ``Aspects
of (2,0) String Compactifications,'' {\it Nucl. Phys.} {\bf B304} (1988) 1
\semi B. Greene, ``Superconformal Compactifications in Weighted Projective
Space,'' {\it Comm. Math. Phys.} {\bf 130} (1990) 335.}
\lref\bagger{E.~ Witten and J. Bagger, Phys. Lett. {\bf 115B}(1982) 202}
\lref\witcurrent{E.~ Witten,``Global Aspects of Current Algebra'',
Nucl.Phys.B223 (1983) 422\semi ``Current Algebra, Baryons and Quark
Confinement'', Nucl.Phys. B223 (1993) 433}
\lref\Wittreiman{S.B. Treiman, E. Witten, R. Jackiw, B. Zumino, ``Current
Algebra and Anomalies'', Singapore, Singapore: World Scientific ( 1985) }
\lref\Witgravanom{L. Alvarez-Gaume, E.~ Witten, ``Gravitational Anomalies'',
Nucl.Phys.B234:269,1984. }

\lref\nicolai{\paper {H.~Nicolai}{New Linear Systems for 2D Poincar\'e
Supergravities}{\NPB{414}{1994}{299},}{\hh 9309052}.}

%%%%%%
%% References herein
%%%%%%%%%%%%%%%%%%%%%%%%%%%%%%%%%%%%%%%%%%%%%%%%

%%\lref\NAME{\paper
%%{Authors}{Title(in \sl)}{\PLB{No.}{Year}{page},}
%%{\hh 0008248 (in\tt)}.}

\lref\bg{\paper {L.~Baulieu, B.~Grossman}{Monopoles and Topological Field
Theory}{\PLB{214}{1988}{223}.}{}}

\lref\seibergsix{\paper {N.~Seiberg}{Non-trivial Fixed Points of The
Renormalization Group in Six
 Dimensions}{\PLB{390}{1997}{169}}{\hh 9609161}\semi
\paper {O.J.~Ganor, D.R.~Morrison, N.~Seiberg}{
  Branes, Calabi-Yau Spaces, and Toroidal Compactification of the N=1
  Six-Dimensional $E_8$ Theory}{\NPB{487}{1997}{93}}{\hh 9610251}\semi
\paper {O.~Aharony, M.~Berkooz, N.~Seiberg}{Light-Cone
  Description of (2,0) Superconformal Theories in Six
  Dimensions}{Adv. Theor. Math. Phys. {\bf 2} (1998) 119}{\hh 9712117.}}

\lref\cs{\paper {L.~Baulieu}{Chern-Simons Three-Dimensional and
Yang--Mills-Higgs Two-Dimensional Systems as Four-Dimensional Topological
Quantum Field Theories}{\PLB{232}{1989}{473}.}{}}

\lref\beltrami{\paper {L.~Baulieu, M.~Bellon}{Beltrami Parametrization and
String Theory}{\PLB{196}{1987}{142}}{}\semi
\paper {L.~Baulieu, M.~Bellon, R.~Grimm}{Beltrami Parametrization For
Superstrings}{\PLB{198}{1987}{343}}{}\semi
\paper {R.~Grimm}{Left-Right Decomposition of Two-Dimensional Superspace
Geometry and Its BRS Structure}{Annals Phys. {\bf 200} (1990) 49.}{}}

\lref\bbg{\paper {L.~Baulieu, M.~Bellon, R.~Grimm}{Left-Right Asymmetric
Conformal Anomalies}{\PLB{228}{1989}{325}.}{}}

\lref\bonora{\paper {G.~Bonelli, L.~Bonora, F.~Nesti}{String Interactions
from Matrix String Theory}{\NPB{538}{1999}{100},}{\hh 9807232}\semi
\paper {G.~Bonelli, L.~Bonora, F.~Nesti, A.~Tomasiello}{Matrix String Theory
and its Moduli Space}{}{\hh 9901093.}}

\lref\corrigan{\paper {E.~Corrigan, C.~Devchand, D.B.~Fairlie,
J.~Nuyts}{First Order Equations for Gauge Fields in Spaces of Dimension
Greater Than Four}{\NPB{214}{452}{1983}.}{}}

\lref\acha{\paper {B.S.~Acharya, M.~O'Loughlin, B.~Spence}{Higher
Dimensional Analogues of Donaldson-Witten Theory}{\NPB{503}{1997}{657},}{\hh
9705138}\semi
\paper {B.S.~Acharya, J.M.~Figueroa-O'Farrill, M.~O'Loughlin,
B.~Spence}{Euclidean
  D-branes and Higher-Dimensional Gauge   Theory}{\NPB{514}{1998}{583},}{\hh
  9707118.}}

\lref\Witr{\paper{E.~Witten}{Introduction to Cohomological Field   Theories}
{Lectures at Workshop on Topological Methods in Physics (Trieste, Italy, Jun
11-25, 1990), \IJMP{A6}{1991}{2775}.}{}}

\lref\ohta{\paper {L.~Baulieu, N.~Ohta}{Worldsheets with Extended
Supersymmetry} {\PLB{391}{1997}{295},}{\hh 9609207}.}

\lref\gravity{\paper {L.~Baulieu}{Transmutation of Pure 2-D Supergravity
Into Topological 2-D Gravity and Other Conformal Theories}
{\PLB{288}{1992}{59},}{\hh 9206019.}}

\lref\wgravity{\paper {L.~Baulieu, M.~Bellon, R.~Grimm}{Some Remarks on  the
Gauging of the Virasoro and   $w_{1+\infty}$
Algebras}{\PLB{260}{1991}{63}.}{}}

\lref\fourd{\paper {E.~Witten}{Topological Quantum Field
Theory}{\CMP{117}{1988}{353}}{}\semi
\paper {L.~Baulieu, I.M.~Singer}{Topological Yang--Mills Symmetry}{Nucl.
Phys. Proc. Suppl. {\bf 15B} (1988) 12.}{}}

\lref\topo{\paper {L.~Baulieu}{On the Symmetries of Topological Quantum Field
Theories}{\IJMP{A10}{1995}{4483},}{\hh 9504015}\semi
\paper {R.~Dijkgraaf, G.~Moore}{Balanced Topological Field
Theories}{\CMP{185}{1997}{411},}{\hh 9608169.}}

\lref\wwgravity{\paper {I.~Bakas} {The Large $N$ Limit   of Extended
Conformal Symmetries}{\PLB{228}{1989}{57}.}{}}

\lref\wwwgravity{\paper {C.M.~Hull}{Lectures on $\CW$-Gravity,
$\CW$-Geometry and
$\CW$-Strings}{}{\hh 9302110}, and~references therein.}

\lref\wvgravity{\paper {A.~Bilal, V.~Fock, I.~Kogan}{On the origin of
$W$-algebras}{\NPB{359}{1991}{635}.}{}}

\lref\surprises{\paper {E.~Witten} {Surprises with Topological Field
Theories} {Lectures given at ``Strings 90'', Texas A\&M, 1990,}{Preprint
IASSNS-HEP-90/37.}}

\lref\stringsone{\paper {L.~Baulieu, M.B.~Green, E.~Rabinovici}{A Unifying
Topological Action for Heterotic and  Type II Superstring  Theories}
{\PLB{386}{1996}{91},}{\hh 9606080.}}

\lref\stringsN{\paper {L.~Baulieu, M.B.~Green, E.~Rabinovici}{Superstrings
from   Theories with $N>1$ World Sheet Supersymmetry}
{\NPB{498}{1997}{119},}{\hh 9611136.}}

\lref\bks{\paper {L.~Baulieu, H.~Kanno, I.~Singer}{Special Quantum Field
Theories in Eight and Other Dimensions}{\CMP{194}{1998}{149},}{\hh
9704167}\semi
\paper {L.~Baulieu, H.~Kanno, I.~Singer}{Cohomological Yang--Mills Theory
  in Eight Dimensions}{ Talk given at APCTP Winter School on Dualities in
String Theory (Sokcho, Korea, February 24-28, 1997),} {\hh 9705127.}}

\lref\witdyn{\paper {P.~Townsend}{The eleven dimensional supermembrane
revisited}{\PLB{350}{1995}{184},}{\hh9501068}\semi
\paper{E.~Witten}{String Theory Dynamics in Various Dimensions}
{\NPB{443}{1995}{85},}{\hh 9503124}.}

\lref\bfss{\paper {T.~Banks, W.Fischler, S.H.~Shenker,
L.~Susskind}{$M$-Theory as a Matrix Model~:
A~Conjecture}{\PRD{55}{1997}{5112},} {\hh9610043.}}

\lref\seiberg{\paper {N.~Seiberg}{Why is the Matrix Model
Correct?}{\PRL{79}{1997}{3577},} {\hh 9710009.}}

\lref\sen{\paper {A.~Sen}{$D0$ Branes on $T^n$ and Matrix Theory}{Adv.
Theor. Math. Phys.~{\bf 2} (1998) 51,} {\hh 9709220.}}

\lref\laroche{\paper {L.~Baulieu, C.~Laroche} {On Generalized Self-Duality
Equations Towards Supersymmetric   Quantum Field Theories Of
Forms}{\MPL{A13}{1998}{1115},}{\hh  9801014.}}

\lref\bsv{\paper {M.~Bershadsky, V.~Sadov, C.~Vafa} {$D$-Branes and
Topological Field Theories}{\NPB{463} {1996}{420},}{\hh9511222.}}

\lref\vafapuzz{\paper {C.~Vafa}{Puzzles at Large N}{}{\hh 9804172.}}

\lref\dvv{\paper {R.~Dijkgraaf, E.~Verlinde, H.~Verlinde} {Matrix String
Theory}{\NPB{500}{1997}{43},} {\hh9703030.}}

\lref\wynter{\paper {T.~Wynter}{Gauge Fields and Interactions in Matrix
String Theory}{\PLB{415}{1997}{349},}{\hh9709029.}}

\lref\kvh{\paper {I.~Kostov, P.~Vanhove}{Matrix String Partition
Functions}{}{\hh9809130.}}

\lref\ikkt{\paper {N.~Ishibashi, H.~Kawai, Y.~Kitazawa, A.~Tsuchiya} {A
Large $N$ Reduced Model as Superstring}{\NPB{498} {1997}{467},}{\hh
9612115.}}

\lref\ss{\paper {S.~Sethi, M.~Stern} {$D$-Brane Bound States
Redux}{\CMP{194}{1998} {675},}{\hh 9705046.}}

\lref\mns{\paper {G.~Moore, N.~Nekrasov, S.~Shatashvili} {$D$-particle Bound
States and Generalized Instantons}{} {\hh 9803265.}}

\lref\bsh{\paper {L.~Baulieu, S.~Shatashvili} {Duality from Topological
Symmetry}{} {\hh 9811198.}}

\lref\pawu{ {G.~Parisi, Y.S.~Wu} {}{ Sci. Sinica  {\bf 24} {(1981)} {484}.}}

%%%%%%%%%%%%%%%%
\lref\coulomb{ {L.~Baulieu, D.~Zwanziger, }   {\it Renormalizable Non-Covariant
Gauges and Coulomb Gauge Limit}, {Nucl.Phys. B {\bf 548 } (1999) 527,} {\hh
9807024}.}

\lref\dan{ {D.~Zwanziger},  {}{Nucl. Phys. B {\bf   139}, (1978) {1}.}{}}

\lref\danzinn{  {J.~Zinn-Justin, D.~Zwanziger, } {}{Nucl. Phys. B  {\bf
295} (1988) {297}.}{}}

\lref\danlau{ {L.~Baulieu, D.~Zwanziger, } {\it Equivalence of Stochastic
Quantization and the Faddeev-Popov Ansatz,
  }{Nucl. Phys. B  {\bf 193 } (1981) {163}.}{}}

\lref\munoz{ { A.~Munoz Sudupe, R. F. Alvarez-Estrada, } {}
Phys. Lett. {\bf 164} (1985) 102; {} {\bf 166B} (1986) 186. }

\lref\okano{ { K.~Okano, } {}
Nucl. Phys. {\bf B289} (1987) 109; {} Prog. Theor. Phys.
suppl. {\bf 111} (1993) 203. }

\lref\singer{
 I.M. Singer, { Comm. of Math. Phys. {\bf 60} (1978) 7.}}

\lref\neu{ {H.~Neuberger,} {Phys. Lett. B {\bf 295}
(1987) {337}.}{}}

\lref\testa{ {M.~Testa,} {}{Phys. Lett. B {\bf 429}
(1998) {349}.}{}}

\lref\Martin{ L.~Baulieu and M. Schaden, {\it Gauge Group TQFT and Improved
Perturbative Yang--Mills Theory}, {  Int. Jour. Mod.  Phys. A {\bf  13}
(1998) 985},   hep-th/9601039.}

%%%%%%%%%%%%%%%%%%%%%%%%%%%%%%%%%%%%%%%%%%%%%%%%%%%%%%%%%%%%%%%%%
\lref\baugros{ {L.~Baulieu, B.~Grossman, } {\it A topological Interpretation
of  Stochastic Quantization} {Phys. Lett. B {\bf  212} {(1988)} {351}.}}

\lref\bautop{ {L.~Baulieu}{ \it Stochastic and Topological Field Theories},
{Phys. Lett. B {\bf   232} (1989) {479}}{}; {}{ \it Topological Field Theories
And Gauge Invariance in Stochastic Quantization}, {Int. Jour. Mod.  Phys. A
{\bf  6} (1991) {2793}.}{}}

\lref\samson{ {L.~Baulieu, S.L.~Shatashvili, { \it Duality from Topological
Symmetry}, {JHEP {\bf 9903} (1999) 011, hep-th/9811198.}}}{}

\lref\halpern{ {H.S.~Chan, M.B.~Halpern}{}, {Phys. Rev. D {\bf   33} (1986)
{540}.}}

\lref\yue{ {Yue-Yu}, {Phys. Rev. D {\bf   33} (1989) {540}.}}

\lref\neuberger{ {H.~Neuberger,} {\it Non-perturbative gauge Invariance},
{ Phys. Lett. B {\bf 175} (1986) {69}.}{}}

\lref\huffel{ {P.H.~Daamgard, H. Huffel},  {}{Phys. Rep. {\bf 152} (1987)
{227}.}{}}

\lref\creutz{ {M.~Creutz},  {\it Quarks, Gluons and  Lattices, }  Cambridge
University Press 1983, pp 101-107.}

\lref\zinn{ {J. ~Zinn-Justin, }  {Nucl. Phys. B {\bf  275} (1986) {135}.}}

\lref\shamir{  {Y.~Shamir,  } {\it Lattice Chiral Fermions
  }{ Nucl.  Phys.  Proc.  Suppl.  {\bf } 47 (1996) 212,  hep-lat/9509023;
V.~Furman, Y.~Shamir, Nucl.Phys. B {\bf 439 } (1995), hep-lat/9405004.}}

 \lref\kaplan{ {D.B.~Kaplan, }  {\it A Method for Simulating Chiral
Fermions on the Lattice,} Phys. Lett. B {\bf 288} (1992) 342; {\it Chiral
Fermions on the Lattice,}  {  Nucl. Phys. B, Proc. Suppl.  {\bf 30} (1993)
597.}}

\lref\neubergerr{ {H.~Neuberger, } {\it Chirality on the Lattice},
hep-lat/9808036.}

\lref\zbgr {L.~Baulieu and D. Zwanziger, {\it QCD $_4$ From a
Five-Dimensional Point of View},    hep-th/9909006.}

\lref\vanish{ {D. ~Zwanziger,} {\it Vanishing of zero-momentum lattice gluon
propagator and color confinement,} {Nucl. Phys. B {\bf  364} (1991) {127}.}}

\lref\cuzwsc{ {Attilio Cucchieri, Daniel Zwanziger, } {\it  Static
Color-Coulomb Force, }{Phys. Rev. Lett. 78 (1997) 3814-3817.}}

\lref\ZZ{ {Ismail Zahed, Daniel Zwanziger, } {\it  Zero Color Magnetization in
QCD Matter, }{Phys. Rev. D61 (2000) 037501.}}

%%%%%%%%%%%%%%%%%%%%%%%%%%%%%%%%%%%%%%%%%%%%%%%%%%%%%%%%%%%%%%%%%%%%%%%%%%%

\lref\cuzwns{A. Cucchieri and D. Zwanziger,
                {\it Numerical study of gluon propagator and 
                 confinement scenario in minimal Coulomb gauge},
                hep-lat/0008026.}

\lref\coul{D. Zwanziger,
               {\it Renormalization in the Coulomb gauge and
                    order parameter for confinement in QCD},
               Nucl.\ Phys.\ {\bf  B518} (1998) 237.}

\lref\Szcz{A. Szczepaniak et al.,
           % Eric S. Swanson, Chueng-Ryong Ji, Stephen R.  Cotanch,
               {\it Glueball spectroscopy in a relativistic
                    many-body approach to hadron structure},
               Phys.\ Rev.\ Lett.\ {\bf 76} (1996) 2011.}

\lref\Robertson{D. G. Robertson et al.,
                % E. S. Swanson, A. P. Szczepaniak, C.-R. Ji, S. R. Cotanch,
               {\it  Renormalized effective QCD Hamiltonian: gluonic sector},
               Phys.\ Rev.\ {\bf D59} (1999) 074019.}

\lref\taylor{J. C. Taylor, 
              {\it Physical and non-standard gauges},
              Proc., Vienna, Austria, 1989, ed. P. Gaigg, W. Kummer and M.
               Schweda (Springer, Berlin, 1990) p.\ 137.}

\lref\doust{P. Doust, 
               {\it Perturbative ambiguities in Coulomb gauge QCD},
               Annals Phys.\ {\bf 177} (1987) 169.}

\lref\recoul{L. Baulieu and D. Zwanziger,
               {\it Renormalizable non-covariant gauges and Coulomb
                 gauge limit}, 
               Nucl.\ Phys.\ {\bf B548} (1999) 527.}

\lref\gribov{V. N. Gribov,
               {\it Quantization of non-Abelian gauge theories},
               Nucl.\ Phys.\ {\bf B139} (1978) 1.}

\lref\critical{D. Zwanziger,
               {\it Critical limit of lattice gauge theory},
               Nucl.\ Phys.\ {\bf B378} (1992) 525.}

\lref\tdlee{T. D. Lee, 
              {\it  Particle physics and introduction to field theory},
                Harwood (1981) p.\ 455.}

\lref\pesschro{M. Peskin and D. Schroeder,
                {\it An introduction to field theory},
                Perseus Books (1995) p.\ 593.}

%%%%%%%%%%%%%%%%%%%%%%%%%%%% CAPTIONS %%%%%%%%%%%%%%%%%%%%%%%%%%%%%%%%%%%%%

\nfig\compar{
Horizontal lines correspond to instantaneous propagators $1/\vk^2$.
Curved lines correspond to non-instantaneous propagators
$1/(\vk^2 + k_4^2)$.  Diagram 1a is a contribution to $V_0$.  Diagram
1b is a contribution to $P_0$.}

%%%%%%
%\draft

%%%%%%%%%

\Title{\vbox
{\baselineskip 10pt
\hbox{hep-th/0008248}
\hbox{NYU-TH-PH-20.8.00}
\hbox{BI-TP 2000/30}
 \hbox{   }
}}
{\vbox{\vskip -30 true pt
\centerline{ Renormalization-group Calculation}
\medskip
 \centerline{ of Color-Coulomb Potential  }
\medskip
\vskip4pt }}
\centerline{{\bf Attilio Cucchieri}$^{\star}$\foot{
Address after February 1st, 2001:
IFSC-USP, Caixa Postal 369, 13560-970 S\~ao Carlos, SP, Brazil.}
  and  {\bf  Daniel
Zwanziger}$^{ \ddag}$}
\centerline{attilio@Physik.Uni-Bielefeld.DE, Daniel.Zwanziger@nyu.edu}
\vskip 0.5cm
\centerline{\it $^{\star}$Fakult\"at f\"ur Physik, Universit\"at Bielefeld,
                 D-33615 Bielefeld, GERMANY}

%\centerline{\it $^{\dag}$ ???}
%\centerline{\it $^{\S}$????}
\centerline{\it $^{\ddag}$   Physics Department, New York University,
New-York,  NY 10003,  USA}

\medskip
\vskip  1cm
\noindent

\def\vx{\vec{x}}
\def\vk{\vec{k}}

\def\cl{{\rm coul}}

	We report here on the application of the perturbative
renormalization-group to the Coulomb gauge in QCD.  We use it to
determine the high-momentum asymptotic form of the instantaneous
color-Coulomb potential $V(\vk)$ and of the vacuum polarization $P(\vk,
k_4)$.  These quantities are renormalization-group invariants, in the
sense that they are independent of the renormalization scheme.  A
scheme-independent definition of the running coupling constant
is provided by
$\vk^2 V(\vk) = x_0 g^2(\vk/\L_\cl)$, and of
$\a_s \equiv { {g^2(\vk/\L_\cl)} \over {4\pi}}$,
where $x_0 = { {12N} \over {11N - 2N_f} }$, and $\L_\cl$ is a finite QCD
mass scale.  We also show how to calculate the coefficients in the
expansion of the invariant $\b$-function
$\b(g) \equiv |\vk| { {\p g} \over{\p |\vk|} }
= -(b_0g^3 + b_1g^5 +b_2g^7 + \dots)$,  where {\it all} coefficients are
scheme-independent.

\Date{\ }

\def\demi{{1\over 2}}

\def\a{\alpha}
\def\b{\beta}
\def\d{\delta}
\def\c{\gamma}
\def\m{\mu}
\def\n{\nu}

\def\r{\rho}

\def\l{\lambda}
\def\L{\Lambda}

\def\LQ{\L_{QCD}}

\newsec{Introduction}

		In QCD the Wilson loop is the basic gauge-invariant
observable.  A
rectangular Wilson loop of dimension $R \times T$ has, asymptotically
at large $T$, the form $W(R,T) \sim \exp[-TV_W(R)]$.  The Wilson
potential $V_W(R)$ is the total energy of the state of lowest energy that
contains a  pair of infinitely massive external quarks at separation
$R$.  If dynamical quarks are present in the theory, the pair of
external quarks polarizes the vacuum and extracts a pair of dynamical
quarks from it, so that, for $R$ not too small, a pair of mesons is
formed at separation $R$, each meson being formed of one external quark
and one dynamical quark, plus doses of other constituents.
In this case $V_W(R)$ represents the total energy of the pair of
mesons at separation $R$.  Clearly the Wilson potential $V_W(R)$ is not
a color-confining potential, but rather a residual potential that
remains after color has been saturated by vacuum polarization.  In this
respect $V_W(R)$ should be regarded as a QCD analog of the van der Waals
interatomic potential.  Clearly it is not the van der Walls force that
holds the atom together, but rather the electrostatic Coulomb
potential.  In the present article we shall be concerned with the QCD
analog of the electrostatic Coulomb potential, which we regard as
responsible for confinement of color charge.

	This quantity is the the color-Coulomb potential $V(R)$.  Just as the
electrostatic Coulomb potential is the instantaneous part of the
44-component of the photon propagator in the Coulomb gauge, likewise
$V(R)$ is the instantaneous part of the 44-component
of the gluon propagator in the (minimal) Coulomb gauge, defined in
Eq.\ (1.3) below.  It is not gauge invariant like the Wilson
potential $V_W(R)$, but it is a more elementary quantity, in terms of
which one may hope to understand the dynamics of the underlying theory.
Indeed we expect that its linear rise (or not) at large $R$ may serve as
an order-paramenter for the confinement of color-charge, even when
$V_W(R)$ is not linearly rising due to vacuum polarization.  In the
present article, we report on a renormalization-group calculation of
$V(R)$ in the high-momentum regime, more precisely, a calculation of its
Fourier transform $V(|\vk|)$ at large $|\vk|$.

	To be sure, confinement manifests itself rather at low momentum.  For
information in this region we have turned to numerical study, and
in an accompanying article on SU(2) lattice gauge theory without
quarks \cuzwns, we present a numerical determination of $V(|\vk|)$ and
also of the equal-time would-be physical 3-dimensionally transverse
gluon propagator $D_{ij}^{\rm tr}(\vk)$.  The reader may find further
references in \cuzwns, and a confrontation of the confinement scenario in
the Coulomb gauge with the numerical data.  This confinement scenario is
also discussed in \coul, and in Sect.\ 3 of the present article.

	The Coulomb gauge is a ``physical gauge" in the sense that the
constraints are solved exactly, including in particular Gauss's law,
$D_iE_i = \r_{\rm qu}$, that is essential for color confinement.  In a
confining theory, all physical states are bound states, and the Coulomb
gauge is the preferred gauge for calculations of  bound-states. Binding
is provided by the instantaneous color-Coulomb potential $V(R)$ that is
treated non-perturbatively, while everything else is regarded as a
perturbation \Szcz, \Robertson.  However the Coulomb gauge offers no
particular advantage for purely perturbative calculations.

	In the present article we shall apply the perturbative
renormalization-group in the Coulomb gauge. This presents new
challenges.  However they are amply rewarded by particularly strong
results.  These are a consequence of the fact that in the Coulomb gauge
the 4-component of the vector potential is invariant under
renormalization \coul,
\eqn\basfac{\eqalign{
g_0A_{0,4}(x) = g_rA_{r,4}(x),
}}
where the subscripts $0$ and $r$ refer to unrenormalized and renormalized
quantities.  This is not true in a Lorentz-covariant gauge.

	The Coulomb gauge is traditionally defined by the condition
$\p_iA_i = 0$.  This is an incomplete gauge fixing because gauge
transformations $g(t)$ that depend on $t$ are not fixed, and
consequently in higher-order calculations one encounters singular
expressions whose evaluation is ambiguous \taylor, \doust.
No doubt this has been an
obstacle to the use of the Coulomb gauge in QCD.  These difficulties are
overcome however by defining the Coulomb gauge as the limit of the
interpolating gauge characterized by the condition
$\p_iA_i + \l \p_4 A_4 = 0$, where $\l$ is a real positive
parameter~\recoul.  Calculations are done at finite $\l$, and at the end
one takes the limit $\l \to 0$.  The interpolating gauge is a
renormalizable gauge that may be treated by standard BRST methods.  The
gauge-fixing term breaks Lorentz invariance, but it is BRST-exact.  An
extension of the BRST operator to include infinitesimal Lorentz
transformations assures Lorentz invariance for BRST-invariant
observables~\recoul.  This definition  of the Coulomb gauge is sufficient
for the purposes of perturbation theory, and for the perturbative
renormalization-group that is the subject of the present article.  In
fact the only calculations that we will do explicitly  (in
Appendix B) are one-loop, and for these one may set $\l = 0$ directly.
However in our discussion of the renormalization group we rely on the
existence of the limit $\l \to 0$ in every order of perturbation theory.

	[At the non-perturbative level with, say, lattice
regularization, the interpolating gauge condition is still subject to the
Gribov ambiguity.  This is resolved by the further specification of the
{\it minimal} Coulomb gauge, whose lattice definition is given in
\cuzwns.  Its continuum analog would be to first minimize
$F_{{\rm hor},A}(g) \equiv \int d^3x \ {^g\!A}_i^2$, for each $t$,  with
respect to all local gauge transformations $g(x,t)$, where
${^g\!A}_\m = g^{-1}A_\m g + g^{-1}\p_\m g$  is the gauge transform of
$A$.  At the minimum, the
Coulomb gauge condition $\p_i A_i = 0$ is satisfied for each $t$, and the
Faddeev-Popov operator $M(A) = -D_i(A)\p_i$ is positive.
Next $F_{{\rm ver},A}(g) \equiv \int d^4x \ {^g\!A}_4^2$
is minimized with respect to $t$-dependent gauge transformations
$g(t)$.  At the minimum $\int d^3x A_4^a(\vx, t) = C$, where $C$ is a
constant that satisfies $C \leq 2\pi V/T$.  Here $V$ is the spatial
3-volume and $T$ is the temporal extent of the Euclidean 4-volume.]

	An immediate consequence of \basfac\ is that the 44-component of the
gluon propagator is also invariant under renormalization.  This gives
the relations
\eqn\inv{\eqalign{
D_{44}(\vk, k_4,\LQ)
= D_{0,44}(\vk, k_4, g_0, \L)
= D_{r,44}(\vk, k_4, g_r, \m).
}}
(We use the same symbol to designate a quantity and its Fourier
transform, but distinguish them by their argument $x$ or $k$.)
Here $\LQ$ is a finite QCD scale, $\L$ is the ultraviolet
cut-off, and $\m$ is the renormalization mass.  We have written the
finite quantity $D_{44}(\vk, k_4,\LQ)$ without a subscript 0 or $r$
because it is a renormalization-group invariant in the sense that it is
independent of $\L$ and $\m$ and of the regularization and
renormalization schemes.  The finite scale $\LQ$, for example
$\LQ = \L_{\rm latt}$ or $\LQ = \L_{\bar{ms}}$, appears in
$D_{0,44}(\vk, k_4, g_0, \L)$ only through its dependence on
$g_0 = g_0(\L/\LQ)$, and similarly for
$D_{r,44}(\vk, k_4, g_r, \m)$.  The scale is defined within the
particular regularization scheme that is used for calculations, for
example lattice or dimensional regularization.  To relate two different
scales, one calculates the same renormalization-group invariant in
the two schemes.

	In the Coulomb gauge $D_{44}(\vx, t)$ has the decomposition into an
instantaneous part, proportional to $\d(t)$ and a
non-instantaneous part $P(\vx, t)$ that is less singular at $t = 0$,
\eqn\deca{\eqalign{
D_{44}(\vx, t) = V(\vx)\d(t) + P(\vx, t),
}}
where $V(\vx)$ is what we call the instantaneous color-Coulomb
potential, and $P(\vx, t)$ is a vacuum polarization term.  (See Eqs.\
(2.13)-(2.15) below.)  We shall see that $V(\vx)$ is anti-screening,
whereas $P(\vx, t)$ is screening.  In momentum space  this decomposition
reads
\eqn\decam{\eqalign{
D_{44}(\vk, k_4) = V(\vk) + P(\vk, k_4).
}}
The instantaneous part is independent of $k_4$.  The $k_4$-independent
part is defined, without reference to a diagrammatic expansion, by
\eqn\defv{\eqalign{
V(\vk) \equiv \lim_{k_4 \to \pm\infty}D_{44}(\vk, k_4).
}}
Here it is assumed that the limit \defv\ exists and is independent of the
sign of $k_4$.  This implies that
\eqn\ninstp{\eqalign{
P(\vk, k_4) \equiv D_{44}(\vk, k_4) - V(\vk).
}}
vanishes at large $k_4$,
\eqn\vatlk{\eqalign{
\lim_{k_4 \to \pm\infty}P(\vk, k_4) = 0.
}}

 The only tool available to calculate these quantities
analytically is perturbation theory.  Perturbatively, in
$d < 4$ dimensions, the separation into instantaneous and vacuum
polarization parts may be made diagram by diagram. Each diagram is
either independent of $k_4$ or satisfies \vatlk.  (See the
decomposition (2.13) -- (2.15) below.)  However in $d = 4$ dimensions,
individual vacuum polarization diagrams contain powers of $\ln(\L k_4)$
associated with divergences, and the limit \defv\ does not exist in any
finite order of perturbation theory.  As a result there is an ambiguity
in the separation of the instantaneous and vacuum polarization parts
in any finite order of perturbation theory.  (This may be regarded as a
mixing of $V$ and $P$ due to renormalization.)   However because $D_{44}$
is a renormalization-group invariant, it follows from the above
definition that $V(\vk)$ and $P(\vk, k_4)$ are separately
renormalization-group invariants.  This will allow us to resolve the
ambiguity by use of the perturbative renormalization-group.  We shall
then find the asymptotic form of $V(\vk)$ at large $|\vk|$ and of
$P(\vk, k_4)$ at large $k_4$.

	As a result of a renormalization-group calculation, we find
that in an SU(N) gauge theory with $N_f$ quarks in the
fundamental representation, the instantaneous color-Coulomb potential
is given by
\eqn\basfor{\eqalign{
\vk^2 V(\vk) = x_0 g^2(|\vk|/\L_\cl),
}}
where
\eqn\xoo{\eqalign{
x_0 = { {12N} \over {11N - 2N_f}},
}}
and $\L_\cl$ is a finite QCD mass scale specified below.  The running
coupling constant $g(|\vk|/\L_\cl)$ is found as the solution of the
renormalization-group equation
\eqn\fleq{\eqalign{
 |\vk| { {\p g} \over {\p |\vk|} }  = \b(g),
}}
where the $\b$-function has the weak-coupling expansion
\eqn\ceb{\eqalign{
\b(g) = - (b_0 g^3 + b_1 g^5 + b_2 g^7 + \dots ),
}}
where {\it all} coefficients $b_n$ are scheme-independent, not
just $b_0$ and $b_1$.  The coefficients
$b_0$ and $b_1$ are also gauge-invariant and have their standard values.
We shall show how to calculate the remaining coefficients
perturbatively in the Coulomb gauge.  This allows a calculation of
$V(\vk)$ to arbitrary accuracy in the high-momentum region.  Its leading
asymptotic behavior is given by
\eqn\lasb{\eqalign{
\vk^2 V(|\vk|)
\sim { {x_0} \over { \ 2b_0 \ln(|\vk|/\L_\cl)} }.
}}
We suppose that asymptotic freedom holds, which requires that $b_0$,
given explicitly below, be positive, $b_0 > 0$.  This is equivalent to
$11N - 2N_f > 0$.  We observe that $x_0 > 0$, and in fact
$x_0 > 1$.

	We have also found the leading asymptotic behavior of
the vacuum polarization term $P(\vk, k_4)$ at large $k_4$ namely,
\eqn\nlcl{\eqalign{
\vk^2 P^{\rm as}(|\vk|, k_4, g_0, \L)
\sim { {y_0} \over { \ 2b_0 \ln(k_4/\L_\cl')} },
}}
where $\L_\cl'$ is another finite QCD mass scale, and
$y_0 = 1 - x_0 = - \ { {N + 2N_f} \over {11N - 2N_f} } < 0$.
The negative sign of $y_0$ shows that the vacuum polarization term is
indeed a screening term.

\newsec{Formula for $D_{44}$ in the Coulomb gauge}

	In the minimal Coulomb gauge, the Euclidean partition function is
given by the
familiar Faddeev-Popov formula
\eqn\part{\eqalign{
Z(\vec{J}, J_4) \equiv \int_G d^4A \
\d(\p_iA_i) \det(- D_i\p_i)  \exp \int d^4x  \
[-(1/4)F_{\m\n}^2 - i g_0J_\m A_\m)],
}}
where $F_{\m \n}$ is the Yang-Mills field.  We have introduced sources
$J_\m$, in terms of which the gluon propagator is given by
\eqn\gen{\eqalign{
D_{0,\m\n}(x-y) \d^{ab}
\equiv g_0^2\langle A_\m^a(x) A_\n^b(y) \rangle
= -  Z^{-1}{{\d} \over {\d J_\m^a(x)}}
{{\d Z} \over {\d J_\n^b(y)}}|_{J = 0}.
}}
(We shall frequently suppress color indices.)
The subscript $G$ on the integral refers to the fact that the integral over
$A_i^{\rm tr}$ is restricted to within the Gribov region.  This has very
important
dynamical consequences at long range that have been proposed as a mechanism
for the
confinement of color charge \gribov\ and \critical, and which are
substantiated by
numerical studies of the Coulomb gauge \cuzwns.   However in the present
article we are interested in the asymptotically high-momentum region
where asymptotic freedom reigns, and we need not specify the Gribov
region $G$.

	The partition function may also be expressed in terms of the
first-order or
phase-space functional integral, by introducing a Gaussian integral over
an independent color-electric field $E_i^a$.
This is done in Appendix A, with the result that, in the
minimal Coulomb gauge, the partition function may be expressed as an integral
over the physical, canonical degrees of freedom $A_i^{\rm tr}$ and $E_i^{\rm
tr}$ with the canonical action
\eqn\pcoul{\eqalign{
Z(\vec{J}, J_4) = \int_G & dA^{\rm tr} dE^{\rm tr} \     \cr
& \times \exp
  \int d^4x
\ ( iE_i^{\rm tr}\dot{A}_i^{\rm tr} - {\cal H} - i g_0 J_i A_i^{\rm tr} ).
}}
Here
\eqn\energy{\eqalign{
  {\cal H}  = (1/2) (E_i^2 + B_i^2)
}}
is the classical Hamiltonian density,
$B_1 \equiv \p_2 A_3 - \p_3 A_2 + g_0 A_2 \times A_3$, etc.,
where $A_i^a = A_i^{{\rm tr},a}$,
$(X \times Y)^a \equiv f^{abc} X^b Y^c$, and the color-electric field
$E_i$ is expressed in terms of the canonical variables by solving Gauss's
law,
\eqn\gslaw{\eqalign{
D_iE_i = g_0 J_4,
}}
where $D_i \equiv \p_i + g_0A_i \times$ is the gauge-covariant
derivative.

	Gauss's law is solved by separating the transverse and longitudinal
components of $E_i$ according to
\eqn\colore{\eqalign{
E_i = E_i^{\rm tr} - \p_i \f.
}}
Here $\f$ is the color-Coulomb potential in terms of which Gauss's law
reads
$-D_i\p_i\f + g_0A_i^{\rm tr} \times E_i^{\rm tr} = g_0J_4$, or
\eqn\gslawa{\eqalign{
M(A^{\rm tr}) \f = \r_{\rm coul} + g_0 J_4,
}}
where $M(A^{\rm tr}) \equiv - D_i(A^{\rm tr}) \p_i$
 is the 3-dimensional Faddeev-Popov operator, and
$\r_{\rm coul} \equiv  - g_0 A_i^{\rm tr} \times E_i^{\rm tr}$
is the color-charge density of the dynamical degrees of freedom.
If we had included dynamical quarks then, in addition to a Dirac term
$\bar{q}(\c_i D_i + m) q$ in the Hamiltonian density, there
would also be a quark contribution to the color-charge density
\eqn\quch{\eqalign{
\r_\cl^a = \r_{\rm gl}^a + \r_{\rm qu}^a =
-g_0 f^{abc} A_i^{{\rm tr},b} E_i^{{\rm tr},c}
+ g_0 \bar{q} \c_0 t^a q.}}
The solution of Gauss's law is given by
\eqn\ssgauss{\eqalign{
\f(\vx, x_4) = \int d^3y
\ M^{-1}[\vx, \vec{y}, A^{\rm tr}(x_4)]
\  (\r_{\rm coul} + g_0 J_4)(\vec{y}, x_4).
}}
Because $M(A^{\rm tr})$
involves only spatial derivatives, the inverse operator $M^{-1}(A^{\rm
tr})$ acts
instantaneously in time.   The Hamiltonian is given by
\eqn\ham{\eqalign{
H & = \int d^3x \ {\cal H} = (1/2) \int d^3x \ (E_i^2 + B_i^2)   \cr
  & = (1/2) \int d^3x \ [E_i^{{\rm tr}2} + (\p_i\f)^2 + B_i^2]     \cr
          & = (1/2) \int d^3x \ (E_i^{{\rm tr}2} + B_i^2)     \cr
   & \ \ + (1/2) \int d^3x \ d^3y \ (\r_{\rm coul} + g_0 J_4)(\vx)
\ {\cal V}(\vx,\vec{y}; A^{\rm tr})
\ (\r_{\rm coul} + g_0 J_4)(\vec{y}).
}}
Here
\eqn\pot{\eqalign{
{\cal V}(\vx, \vec{y}; A^{\rm tr}) \equiv
    [M(A^{\rm tr})^{-1}(-\p^2)M(A^{\rm tr})^{-1}]_{\vec{x},\vec{y}}
}}
is a color-Coulomb potential-energy functional, that depends on
$A^{\rm tr}$.  It acts instantaneously, and couples universally to
color charge including, in the present case, the source~$J_4$.  Its
physical significance will be discussed in the next section.

	We now come to the important point.  We use $Z$ from \pcoul\ and $H$
from \ham, and obtain
\eqn\fndr{\eqalign{
Z^{-1}{ {\d Z}\over {\d J_4^a(x)} } = - \ g_0 \int d^3z \ \langle \
{\cal V}^{ac}(\vx,\vec{z}; A^{\rm tr})
\ [\r_{\rm coul}^c(\vec{z}, x_4) + g_0 J_4^c(\vec{z}, x_4) ]\
\rangle, }}
To calculate $D_{0,44}(x-y)$ we apply
${ {\d }\over {\d J_4^b(y)} }$ and we obtain two terms,
\eqn\dec{\eqalign{
D_{0,44}(x-y) = V_0(x-y) + P_0(x-y).
}}
The first,
\eqn\inst{\eqalign{
V_0(x-y) \d^{ab} \equiv g_0^2 \langle \
{\cal V}^{ab}[\vx, \vec{y}; A^{\rm tr}(x_4)] \ \rangle \ \d(x_4 - y_4),
}}
comes from the $J_4^c(\vec{z}, x_4)$ that appears explicitly in \fndr.
It is manifestly instantaneous. The second term,
\eqn\nst{\eqalign{
P_0(x-y)\d^{ab} \equiv - g_0^2 \langle \  ({\cal V}\r_{\rm coul})^a(x)
\  ({\cal V}\r_{\rm coul})^b(y) \ \rangle,
}}
where
$ ({\cal V}\r)^a(x) = ({\cal V}\r)^a(\vx, x_4)
\equiv \int d^3z \ {\cal V}^{ab}[\vx, \vec{z}; A^{\rm tr}(x_4)]
\r^b(\vec{z},x_4)$, represents polarization of the vacuum.  The
expansion of $V_0$ and $P_0$ up to one-loop is given in Appendix B. The
significance of these two terms for confinement is discussed in the next
section.

\newsec{Physical Interpretation}

	We comment briefly on the physical meaning of the decomposition
\dec\ to \nst.
We showed in \cuzwns\ that the Faddeev-Popov operator $M(A^{\rm tr})$
is a positive operator in the minimal Coulomb gauge.  It follows that
${\cal V}(A^{\rm tr}) = M^{-1}(-\p^2)M^{-1}$ is also a
positive operator.  The $P_0$ term represents ordinary vacuum
polarization that also occurs in QED.  The minus sign that
appears in front of $P_0$ indicates that it corresponds to
screening. In QED, the color-Coulomb potential energy functional
${\cal V}(\vx, \vec{y}, A^{\rm tr})$
would be replaced by the electrostatic Coulomb potential
$(4 \pi |\vx - \vec{y}|)^{-1}$.   In \cuzwns\ we argued that
${\cal V}(A^{\rm tr})$  is long-range for a typical configuration
$A^{\rm tr}$, and presented numerical evidence that this is true.

	According to the confinement scenario in the minimal Coulomb gauge,
${\cal V}(A^{\rm tr})$ is predominantly long range because
$M^{-1}(A^{\rm tr})$ is long range.  This happens because the Gribov
region is bounded by $M(A^{\rm tr}) \geq 0$, and entropy favors dense
population close to the boundary.  The boundary occurs where
$M(A^{\rm tr})$ has a zero eigenvalue, so it has a small eigenvalue
close to the boundary.  As a result $M^{-1}(A^{\rm tr})$
is enhanced close to the boundary where the population is dense.  The
same color-Coulomb potential energy functional
${\cal V}(A^{\rm tr})$ appears in both the instantaneous term $V_0$ and
the vacuum polarization term $P_0$, so both are long range.

	The instantaneous term $V_0$ is responsible for confinement,
whereas the
vacuum polarization term $P_0$ causes screening.  In any physical process
both $V_0$ and $P_0$ contribute of course, and one or the other may
dominate.  Consider the Wilson loop which is a model of a pair of
infinitely heavy external quarks.  In a theory of pure glue without
dynamical quarks, it is believed that at long range there is a linearly
rising potential $V_W(R)$ in the Wilson loop, characterized by a string
tension.   In this case, according to the confinement scenario in the
Coulomb gauge, the instantaneous term $V_0$ dominates.  However if
dynamical quarks are present then it is believed that the string
``breaks" at some distance because it is energetically favorable to
produce pairs of dynamical quarks from the vacuum.  In this case the
vacuum polarization term $P_0$ dominates.  However in both cases
color-charge itself is confined.  According to the confinement scenario
in the Coulomb gauge, that is because the instantaneous term
$V_0(R)$ is long range and presumably linearly rising
even when $V_W(R)$ is not.  Indeed it is precisely the long range of
$V_0(R)$ that makes it energetically favorable to produce dynamical
quark pairs from the vacuum, thereby causing color confinement.  Thus in
the Coulomb gauge the instantaneous part provides an order-parameter for
the confinement of color charge even when dynamical quarks are present.

	As noted, the term $V_0$ is instantaneous.  	It is easy to see
that in
each order of perturbation theory $V_0(x-y)$ is the sum
of diagrams in which the points $x$ and $y$ are continuously connected by
instantaneous free gluon propagators, as illustrated in Fig.\ 1a.
We now show that the diagrams contributing to $P_0(x-y)$ have no
instantaneous parts. Indeed the free propagators of $A^{\rm tr}$ and
$E^{\rm tr}$ are given by
\eqn\prop{\eqalign{ \langle A_i^{\rm tr} A_j^{\rm tr}\rangle_0
	 & =  P_{ij}(\hat{k})/(\vk^2 + k_4^2)       \cr
    \langle E_i^{\rm tr} E_j^{\rm tr}\rangle_0
	 & =  (\vk^2 \d_{ij} - k_i k_j)/(\vk^2 + k_4^2)     \cr
  \langle E_i^{\rm tr} A_j^{\rm tr}\rangle_0
	 & =  P_{ij}(\hat{k})k_4/(\vk^2 + k_4^2)
, }}
where
$P_{ij}(\hat{k}) = (\d_{ij} - \hat{k_i} \hat{k_j})$ is the
3-dimensionally transverse projector.  These propagators vanish in the
limit $k_4 \to \infty$, as do the quark free propagators. From
the structure of the vacuum polarization term \nst,
as illustrated in Fig.\ 1b, one sees that the diagrams contributing to it
must somewhere be connected by the dynamical propagators \prop, so
$P_0$ has no instantaneous part.\foot{It is
essential to use the phase-space integral with first-order action to
separate out the instantaneous diagrams in this way.  For one could also
integrate out
$A_4$ in the second-order Lagrangian formalism.  However if one did that,
then $\dot{A}_i^{\rm tr}$ would appear at vertices instead of the
independent field variable $E_i^{\rm tr}$.  The free propagator
$\langle \dot{A}_i^{\rm tr}\dot{A}_j^{\rm tr}\rangle_0
= P_{ij}(\hat{k})k_4^2/(\vk^2 + k_4^2)$ does not vanish in the limit
$k_4 \to \infty$, but rather has the limit
$\langle \dot{A}_i^{\rm tr}\dot{A}_j^{\rm tr}\rangle_0 \to P_{ij}(\hat{k})$
which has an instantaneous non-local part.}  This is apparent in the
one-loop calculation presented in Appendix B.

\newsec{A technical difficulty}

	In Eq.\ \defv\ we have given a definition of the color-Coulomb potential
which is independent of perturbation theory. However since its defining
property is false in any finite order or perturbation theory, we must
address the problem of how to calculate it.

	The perturbative expansion of
$D_{0,44} = V_0 + P_0$ begins with
\eqn\pe{\eqalign{
V_0(|\vk|, g_0, \L)
     & = g_0^2 \ (\vk^2)^{-1}  + O(g_0^4) \cr
P_0(|\vk|, k_4, g_0, \L)
     & = O(g_0^4).}}
The Feynman integrals for these quantities are derived to one loop
order in Appendix B.  The integral for
$P_0(|\vk|, k_4, g_0, \L)$ is relatively complicated, however it
simplifies considerably for $k_4 >> |\vk|$, as discussed in Appendix B,
and one obtains to one-loop order for SU(N) gauge theory,
\eqn\peo{\eqalign{
\vk^2 V_0(|\vk|, g_0, \L)
     & = g_0^2  + g_0^4[v_{11}\ln(\L/|\vk|) + v_{10}] \cr
\vk^2 P_0(|\vk|, k_4, g_0, \L)
     & = g_0^4[p_{11}\ln(\L/k_4) + p_{10} + o(|\vk|/k_4)],}}
where $\L$ is an ultra-violet cut-off,
$\lim_{x \to 0}o(x) = 0$, and $v_{nm}$ and $p_{nm}$ are constants, with
\eqn\cmp{\eqalign{
v_{11} =  { {8N}\over {(4\pi)^2} }
}}
\eqn\olr{\eqalign{
p_{11} = - { {2} \over {(4\pi)^2} }
\Big( { {1} \over {3} } N +  { {2} \over {3} } N_f \Big).
}}
Here we have included the vacuum polarization contribution of $N_f$
flavors of quark in the fundamental representation which, to this order,
is gauge-independent.  These coefficients are also found in a
Hamiltonian formalism \tdlee.

	We obtain asymptotically for $k_4 >> |\vk|$, for
$D_{0,44}^{\rm as} = V_0 + P_0^{\rm as}$,
\eqn\asy{\eqalign{
\vk^2 D_{0,44}^{\rm as}(|\vk|, k_4, g_0,\L)
= g_0^2 +
g_0^4[v_{11}\ln(\L/|\vk|) + p_{11}\ln(\L/k_4) + v_{10} + p_{10}].
}}
Observe that $p_{10}$ contributes a
$1/\vk^2$ term to $D_{0,44}$.  Consequently we cannot assert that the
instantaneous part of $D_{0,44}$ in $d = 4$ dimensions, namely the
color-Coulomb potential $V$ is given by the sum of instantaneous
diagrams, as it is in $d < 4$ dimensions,
\eqn\ntru{\eqalign{
	V(|\vk|, \LQ) \neq V_0(\vk, g_0, \L).
}}
We shall see that $V_0$ and $P_0$ separately are not
renormalization-group invariants, and this will be the key to an
unambiguous separation of the instantaneous part $V$.  This implies that
the coefficients that appear in $V_0$ and $P_0$ may be scheme
dependent.  In particular, $v_{10}$ and $p_{10}$ are scheme-dependent.

	Another way to see that the separation cannot be made diagram by
diagram is to recall that in higher-order, calculations must be done in
the interpolating gauge, with gauge condition
$\p_i A_i + \l \p_4 A_4 = 0$.  For finite $\l$ no diagram is
instantaneous, and a separation criterion is required for the limiting
expression at $\l = 0$.

	As a first step we note that in one-loop order
$D_{0,44}^{\rm as}(|\vk|, k_4, g_0,\L)$ at large~$k_4$
conveniently separates into a sum of terms that depend respectively on
$|\vk|$ and on $k_4$.  To get the instantaneous part, namely the part
that is independent of $k_4$, we simply delete the term that depends on
$k_4$, namely
$p_{11}\ln(\L/k_4)$.  However the separation of the constant term is
ambiguous, as it is in each order.  We conclude that to one-loop order,
the $k_4$-independent part of
$D_{0,44}(|\vk|, k_4, g_0,\L)$,
 is given by
\eqn\oloi{\eqalign{
\vk^2 V(|\vk|, g_0,\L) = g_0^2 \ x_0 +
g_0^4[v_{11}\ln(\L/|\vk|) + x_1],
}}
where $x_0$ and $x_1$ are as yet unknown constants.

	The form of the higher-order contributions to $P_0$ is not known.
However we shall make the simplest assumption namely that	the same
procedure may be effected to arbitrary order, so that only constant
terms in each order in the expansion of $\vk^2
V(|\vk|, g_0,\L)$ are ambiguous,
\eqn\olof{\eqalign{
\vk^2 V(|\vk|, g_0,\L) = g_0^2 \ x_0 +
\sum_{n=1}^\infty g_0^{2n+2}[x_n +\sum_{m=1}^n v_{nm}\ln^m(\L/|\vk|) ].
}}
Here the coefficients $v_{nm}$ are calculated from the perturbative
expansion of $V_0$, but the $x_n$ are a set of as yet unknown
constants.  (In case there are additional ambiguous terms
in higher order in the separation of the vacuum polarization part $P_0$
from $D_{0,44}$, they may also be determined using the
renormalization-group, because it also restricts the $v_{nm}$.)

	From its definition as the large-$k_4$ limit of $D_{44}$, it follows
that color-Coulomb potential is a renormalization-group invariant.  So
when $V(|\vk|)$ is expressed either in terms of unrenormalized or
renormalized quantities, it is independent of $\L$ and $\m$ (and of the
regularization and renormalization scheme),
\eqn\rgi{\eqalign{
V(|\vk|) \equiv V(|\vk|, g_0(\L/\LQ), \L)
= V(|\vk|, g_r(\m/\LQ), \m).
}}
Consequently we may set
$\L = |\vk|$, and $\m = |\vk|$, which gives
\eqn\rgia{\eqalign{
V(|\vk|) = V(|\vk|, g_0(|\vk|/\LQ), |\vk|)
= V(|\vk|, g_r(|\vk|/\LQ), |\vk|).
}}
(Once the functional dependence of
$D_{0,44}$ or $V_0$ on the cut-off $\L$ is determined -- it is a
polynomial in $\ln
\L$ in each order of perturbation theory -- then $\L$ may be assigned any
finite
value.)
Thus we may set $\L = |\vk|$ in \olof, which gives a simple expansion
in terms of the unknown coefficients $x_n$,
\eqn\olofa{\eqalign{
\vk^2 V(|\vk|) = x_0 \ g_0^2(|\vk|/\LQ)  +
\sum_{n=1}^\infty x_n \ g_0^{2n+2}(|\vk|/\LQ).
}}

\newsec{Renormalization-group in the Coulomb gauge}

	The unrenormalized coupling constant $g_0 = g_0(\L/\LQ)$ is a function
of the cut-off $\L$, determined by the flow equation
\eqn\vco{\eqalign{
\L \ \p g_0 / \p \L = \b_0(g_0),
}}
where $\LQ$ is a finite physical QCD mass scale.
The $\b$-function has the expansion
\eqn\bfz{\eqalign{
\b_0(g_0) = - (b_0 g_0^3 + b_1g_0^5 + b_2g_0^7 + \dots) \  \ .
}}
In general, the coefficients $b_n$ are both scheme and gauge dependent,
except for $b_0$ and $b_1$ that are  scheme and gauge independent.
Similarly the renormalized coupling constant $g_r = g_r(\m/\LQ)$ is a
function of the renormalization mass determined by
\eqn\vcor{\eqalign{
\m \ \p g_r / \p \m = \b_r(g_r).
}}

	In the Coulomb gauge $D_{0,44}$ has no anomalous dimension coming from
multiplicative renormalization.  It therefore obeys the simple
Callan-Symanzik equation
\eqn\cse{\eqalign{[\L \p / \p \L + \b_0(g_0)\p / \p g_0]
D_{0,44}(|\vk|, k_4, g_0, \L) = 0. }}
As a result, in the Coulomb gauge, the $\b$-function may be obtained from
the propagator $D_{0,44}$ -- in fact only $D_{0,44}^{\rm as}$ is
needed -- whereas in covariant gauges a calculation of the vertex
function is necessary.  Indeed from the last equation we have
\eqn\cbf{\eqalign{
\b_0(g_0) = - { { \L \p D_{0,44}/ \p \L  } \over
{ \p D_{0,44}/ \p g_0  } },
}}
where $D_{0,44} = D_{0,44}(|\vk|, k_4, g_0, \L)$.
This holds identically for all values of $|\vk|$, $k_4$ and $\L$, so we
may set $k_4$ to an asymptotically large value and obtain
\eqn\cbf{\eqalign{
\b_0(g_0) = - { { \L \p D_{0,44}^{\rm as}/ \p \L  } \over
{ \p D_{0,44}^{\rm as}/ \p g_0  } }.
}} From the one-loop expression \asy\ for $D_{0,44}^{\rm as}$, we obtain
the first coefficient of the $\b$-function
\eqn\olb{\eqalign{
b_0 = \demi (v_{11} + p_{11}).
}} From \cmp\ and \olr\ we obtain the standard expression
\eqn\bfz{\eqalign{
b_0 =  { {1} \over {(4\pi)^2} }
\Big( { {11} \over {3} } N -  { {2} \over {3} } N_f \Big),
}}
without calculating any vertex function.  All coefficients $b_n$ may be
calculated in this way.

	We take the large $k_4$ limit of the Callen-Symanzik equation \cse, and
observe that
$V(|\vk|, g_0, \L)
= \lim_{k_4 \to \infty}D_{0,44}(|\vk|, k_4, g_0, \L)$
satisfies the same Callan-Symanzik equation,
\eqn\csei{\eqalign{
[\L \p / \p \L + \b_0(g_0)\p / \p g_0] V(|\vk|, g_0, \L) = 0.
}}
Since $g_0 = g_0(\L/\LQ)$
is a solution of the flow equation \vco, the Callan-Symanzik equation
yields
\eqn\dep{\eqalign{
{{d V(|\vk|, g_0(\L/\LQ), \L)} \over {d \L}} = 0, }}
which is again the statement that
$V(|\vk|, g_0(\L/\LQ), \L)$ is independent of $\L$.

\newsec{Renormalization-group to the rescue}

	We shall use the renormalization-group and our knowledge of the
$\b$-function to determine the unknown constants $x_n$.  From the
Callan-Symanzik equation \csei\ for
$V$ we have
\eqn\bet{\eqalign{
\  \ \ \ \p V(|\vk|, g_0, \L)/ \p g_0
 = {{- \L \p  V(|\vk|, g_0, \L)/ \p \L}\over
{\b_0(g_0) } },
}}
This is an identity that holds for all $\L$, and we may simplify
it by setting
$\L = |\vk|$,
\eqn\bet{\eqalign{
\  \ \ \ \p V(|\vk|, g_0, \L)/ \p g_0|_{\L = |\vk|}
 = {{- \L \p  V(|\vk|, g_0, \L)/ \p \L|_{\L = |\vk|}}\over
{\b_0(g_0) } }.
}}
To find all the $x_n$ we substitute the expansion \olof\ on the left
and right hand sides.  We also expand
\eqn\ibet{\eqalign{
\b_0^{-1}(g_0) = - b_0^{-1} g_0^{-3}(1 + \sum_{p=1}^\infty c_pg_0^{2p}),
}}
where, from \bfz, we have $c_1 = -b_1/b_0$ etc.  The derivative with
respect to $\L$ on the right-hand side of \bet\ kills the constants
$x_n$, and we have
\eqn\vte{\eqalign{
\sum_{n=0}^\infty 2(n+1)g_0^{2n+1}x_n
= b_0^{-1} g_0^{-3}\sum_{p=0}^\infty c_pg_0^{2p}
 \sum_{m=1}^\infty g_0^{2m+2}v_{m1},
}}
where $c_0 \equiv 1$.  Equating like powers of $g_0$ we obtain
\eqn\solvx{\eqalign{
x_n = [2(n+1)b_0]^{-1}
\sum_{m=1}^{n+1} c_{n-m+1} v_{m1},
}}
and in particular
\eqn\xow{\eqalign{
x_0 & =  (2b_0)^{-1} v_{11}  \cr
x_1 & = (4b_0)^{-1}(v_{11}c_1 + v_{21}). }}
Thus $x_0$ is found from $v_{11}$ and $b_0$ which require 1-loop
calculations, and $x_1$ requires 2-loop calculations.   Usually the
renormalization-group is used to determine higher-order logarithms from
lower order terms.  Here instead we have used it to consistently
determine an unknown lower-order constant from a known higher-order
logarithm.  From \cmp\ and \bfz, we obtain
\eqn\xoo{\eqalign{
x_0  = { {12N} \over {11N - 2N_f} }.
}}

	It will be convenient in the following to factorize the coefficient
$x_0$ out  of the expansion \olofa\ for $V$, and we write
\eqn\oloh{\eqalign{
\vk^2 V(|\vk|) = x_0 [ \ g_0^2(|\vk|/\LQ) \  +
\sum_{n=1}^\infty x_n' \ g_0^{2n+2}(|\vk|/\LQ) \ ].
}}
where $x_n' \equiv x_n/x_0$.  Remarkably, the leading term is not simply
$g_0^2$, as one would expect from \peo, but rather
$x_0g_0^2 = { {12N} \over {11N - 2N_f} }g_0^2$.

	The leading asymptotic form of $V$ may be found from the expression
\olof\ to order $g_0^4$, with $v_{11} = 2 b_0 x_0$, namely
\eqn\aso{\eqalign{
\vk^2 V(|\vk|) = x_0 g_0^2
\{ \ 1 + g_0^2 [2b_0 \ln(\L/|\vk|) + x_1'] \},
}}
which to this order may be written
\eqn\tho{\eqalign{
\vk^2 V(|\vk|) = x_0
\{ \ g_0^{-2} - [2b_0 \ln(\L/|\vk|) + x_1'] \ \}^{-1},
}}
where
\eqn\xpr{\eqalign{
x_1' = { {x_1} \over {x_0} }
= \demi \big( { {v_{21}} \over{v_{11}} } - { {b_1} \over{b_0} } \big).
}} From the asymptotic form of $g_0^{-2} \sim 2b_0 \ln(\L/\LQ)$, this gives
\eqn\vlk{\eqalign{
\vk^2 V(|\vk|) \sim x_0
[ \ 2b_0 \ln(|\vk|/\LQ) - x_1']^{-1}
}}
\eqn\lcl{\eqalign{
\vk^2 V(|\vk|) \sim { {x_0} \over
{ \ 2b_0 \ln(|\vk|/\L_\cl)} },
}}
which is valid for large $|\vk|$.
Here we have introduced the new mass scale $\L_\cl$ characteristic of the
Coulomb
gauge.  It is related to the scale $\LQ$ used in the scheme
by which the $\b$-function was defined according to
\eqn\nms{\eqalign{
\L_\cl \equiv \exp\big( { {x_1'} \over {2b_0} } \big) \ \LQ .
}}
Normally, a ratio such as $\L_\cl/\LQ$ can be
determined from a 1-loop calculation, but here a 2-loop calculation
of $v_{21}$, the coefficient of a 2-loop logarithm that
appears in Eq.\ \xpr\ would be required.  For just as  the determination
of the constant $x_0$, of 0-loop order, requires the 1-loop calculation
which we report here, similarly, to determine the constant $x'_1$, of
1-loop order, would require a 2-loop calculation.  In both cases the
reason is the ambiguity in the separation of the instantaneous part $V$
which, as noted above, cannot be determined from individual perturbative
diagrams.  Instead it requires the perturbative renormalization group to
determine the n-th order constant from the $(n+1)$-th order logarithm.

	Expression \lcl\ for $V(|\vk|)$ exhibits asymptotic freedom and,
with $x_0$ positive, indeed $x_0 > 1$, the instantaneous part of
$D_{44}$ is anti-screening.

\newsec{Asymptotic form of the non-instantaneous part}

	We may also determine to this order the asymptotic form of the
non-instantaneous part $P = D_{0,44} - V$.  It is also a
renormalization-group invariant.  If we set $k_4$ to an asymptotically
high value we obtain
\eqn\aninstp{\eqalign{
P^{\rm as}(|\vk|, k_4, g_0, \L) \equiv
D_{0,44}^{\rm as}(|\vk|, k_4, g_0, \L) - V(|\vk|, g_0, \L).
}} From \asy\ and \oloi\ this gives to one-loop order
\eqn\nolo{\eqalign{
\vk^2P^{\rm as}(|\vk|, k_4, g_0, \L) =
g_0^2 y_0 + g_0^4[p_{11} \ln(\L/k_4) + y_1],
}}
where, by \olr, \olb\ and \xow,
\eqn\yoyo{\eqalign{y_0 & = 1 - x_0 = (2b_0)^{-1}(2b_0 - v_{11}) =
(2b_0)^{-1}p_{11}   \cr
    & = - \ { {N + 2N_f}\over {11N - 2N_f} },
}}
and $y_1 = p_{10} + v_{10} - x_1$.  This gives
\eqn\noas{\eqalign{
\vk^2 P^{\rm as}(|\vk|, k_4, g_0, \L) =
y_0 g_0^2 \ \{1 + g_0^2[2b_0 \ln(\L/k_4) + y_1'\},
}}
where $y_1' = y_1/y_0$.  By the reasoning that leads to \lcl,
we obtain to this order the asymptotic expression
\eqn\nlcl{\eqalign{
\vk^2 P^{\rm as}(|\vk|, k_4, g_0, \L)
\sim { {y_0} \over { \ 2b_0 \ln(k_4/\LQ')} },
}}
where $\LQ'$ is another finite QCD mass scale like $\L_{\rm coul}$.
It would take also take a 2-loop calculation to determine
$\LQ'/\LQ$.  Since $y_0 < 0 $ is
negative, $P$ is indeed screening.

\newsec{Invariant color charge}

	In this section we show how to calculate $V(|\vk|)$ to
arbitrary accuracy in the ultra-violet region.  We
define a new running coupling by
constant $g = g(|\vk|/\L_\cl)$ by
\eqn\ncc{\eqalign{
\vk^2 V(|\vk|) \equiv x_0 \ g^2(|\vk|/\L_\cl).
}}
Because $V(|\vk|)$ is scheme-independent, $g(|\vk|/\L_\cl)$ is also. From
\oloh\ we have
\eqn\dncc{\eqalign{
g^2(|\vk|/\L_\cl)
= ( \ g_0^2 + \sum_{n=1}^\infty g_0^{2n+2}x_n' \ )|_{g_0 =
g_0(|\vk|/\LQ)},
}}
so the new coupling constant agrees with $g_0$ in lowest order.
Thus it is a regular redefinition of the coupling constant, and
$g$ may be used for perturbative expansions.

	Corresponding to the invariant charge is an invariant
$\b$-function defined by
\eqn\fleq{\eqalign{
\b(g)
\equiv |\vk|{{\p g(|\vk|/\L_\cl) } \over {\p |\vk|} }.
}}
It may be calculated from
\eqn\gleq{\eqalign{
\b(g)
& = {{\p g } \over {\p g_0 } }
|\vk|{{\p g_0(|\vk|/\LQ) } \over {\p |\vk|} } \cr
\b(g)
& =    {{\p g } \over {\p g_0 } }
\ \b_0(g_0)|_{g_0(g)},
}}
where $g_0(g)$ is obtained by inverting Eq.\ \dncc.  This may be
simplified by using
\eqn\ceb{\eqalign{
\b(g)
& =   {{1 } \over {2 g } }
{{\p g^2 } \over {\p g_0 } } \ \b_0(g_0) \cr
 & =   {{\vk^2 } \over {2 gx_0 } }
{{\p V_0 } \over {\p g_0 } }|_{|\vk| = \L}  \ \b_0(g_0) \cr
 & = - {{\vk^2 } \over {2 gx_0 } }
  {{\L \p V_0 } \over {\p \L } }|_{|\vk| = \L},
}}
by Eq.\ \bet.  From Eq.\ \olof, this gives
\eqn\exb{\eqalign{
\b(g) = - { {1} \over { 2g } }
\big( \  2b_0 g_0^4 + \sum_{n=2}^\infty v_{n1}'g_0^{2n+2}\ \big)
|_{g_0(g)},
}}
where $v_{n1}' \equiv v_{n1}/x_0$, and we have used $v_{11}/x_0 = 2b_0$.
To find $V(|\vk|)$ to arbitrary accuracy, one calculates $\b(g)$
perturbatively and then solves the flow equation \fleq\ for
$g(|\vk|/\L_\cl)$.

	Finally we remark that	we may choose new unrenormalized and
renormalized expansion parameters according to
\eqn\newgs{\eqalign{ g_0' & = g(\L/\L_\cl) \cr
g_r' & = g(\m/\L_\cl).}}
so $g_0'$ and $g_r'$ lie on the same invariant trajectory, the only
difference between
them being the value of the argument.

\newsec{Conclusion}

	We have successfully applied the perturbative renormalization group to
the Coulomb gauge.  In this gauge, the 44-component of the gluon
propagator
$D_{44}(|\vk|, k_4, \LQ)$ is  a renormalization-group invariant in the
sense that it is independent of the regularization and renormalization
schemes, and of the ultra-violet cut-off $\L$ and renormalization mass
$\m$.  With the help of the perturbative renormalization-group we have
decomposed it into an instantaneous part $V(|\vk|)$, which we call the
color-Coulomb potential, and a vacuum polarization part $P(|\vk|, k_4)$
which vanishes at large $k_4$.  Each of these terms is separately a
renormalization-group invariant, and their asymptotic form, at large
$|\vk|$ and $k_4$ respectively, was reported in the Introduction.

	The color-Coulomb potential allows us to define an invariant QCD charge
$g(|\vk|/\L_\cl)$ by
$\vk^2 V(\vk) = x_0 g^2(|\vk|/\L_\cl)$,
where $x_0 = { {12N} \over {11N - 2N_f} }$.  This invariant charge is the
QCD analog of the invariant charge of Gell-Mann and Low in QED.  We have
shown how to calculate the corresponding invariant $\b$-function,
$|\vk|{ {\p g} \over {\p |\vk|} }  = \b(g)$.
Because this charge is scheme-independent it may offer some advantage
in providing a definition of
$\a_s(\vk^2) = { {g^2(|\vk|/\L_\cl)} \over {4\pi}}$,
whereas the standard definition in current use is
scheme-dependent (see for example \pesschro). The color-Coulomb
potential $V(|\vk|)$ is also the natural starting point for
calculations of bound-states such as heavy quarkonium.

	The Coulomb gauge provides direct access to quantities
of non-perturbative interest.  Indeed both $V(|\vk|)$ and
$P(|\vk|, k_4)$ have a natural role in a confinement scenario:
$V(|\vk|,\LQ)$ is long-range, anti-screening, and responsible for
confinement of color-charge, whereas the vacuum polarization term
$P(|\vk|, k_4)$ is screening, and responsible for ``breaking of the
string" between external quarks, when dynamical quark pairs are produced
from the vacuum. We expect the linear rise (or not)
of $V(|\vx|)$ at large $|\vx|$ to provide an order parameter for
confinement of color charge, even in the presence of dynamical quarks
when the Wilson loop cannot serve this purpose.
The accompanying article \cuzwns\ reports a numerical study of the
running coupling constant $g^2(\vk/\L_\cl)$.  The data show a
significant enhancement at low $|\vk|$, in agreement with this
confinement scenario.  However additional studies at larger values of $\b
\equiv 4/g_0^2$ are necessary before a conclusion can be reached about a
linear rise of $V(\vx)$  at large $|\vx|$ in the continuum limit, $\b
\to \infty$.

	The data also show a strong suppression of the equal-time,
3-dimensionally transverse, would-be physical, gluon propagator
$D_{ij}^{\rm tr}(\vk)$ at $\vk = 0$, and agree with a formula of
Gribov that {\it vanishes} like $|\vk|$ near $|\vk| = 0$.  The only
explanation for this counter-intuitive behavior is the suppression of
configurations outside the Gribov horizon in the minimal Coulomb gauge.
Since $D_{ij}^{\rm tr}(\vk)$ is strongly suppressed, we may understand
the main long-range forces between color charge as being due to
$D_{44}$ which, as we have seen, is the sum of the attractive
instantaneous color-Coulomb potential $V(|\vk|)$ that is
anti-screening, and the vacuum polarization term $P(|\vk|, k_4)$ that
is screening.  According to the confinement scenario discussed in
Sect.\ 3 and in \coul\ and \cuzwns, both are long-range in the minimal
Coulomb gauge because entropy favors a high density population close to
the Gribov horizon.

\vskip 3mm

  \centerline{\bf Acknowledgments}

The research of Attilio Cucchieri was partially supported by
the TMR network Finite Temperature Phase Transitions in
Particle Physics, EU contract no.: ERBFMRX-CT97-0122.
The research of Daniel Zwanziger was partially supported by the National
Science Foundation under grant PHY-9900769.

\appendix A{Relation of Faddeev-Popov and phase-space functional integrals}

	We wish to derive the canonical or phase-space functional integral
\pcoul\ from the
Faddeev-Popov formula \part.  The argument merely reverses the
text-book derivation of the Faddeev-Popov formula from the canonical
phase-space
functional integral in Coulomb gauge while keeping track of the sources $J_\m$.
We introduce the identity
\eqn\gssid{\eqalign{
\exp[- (1/2) \int d^4x F_{0i}^2 \ ] = N \int d^3E \
\exp \int d^4x [iE_iF_{0i} - (1/2)E_i^2],
}}
which is a Gaussian integral over new variables $E_i^a$ that will
play the role of independent color-electric field variables.
This allows us to rewrite \part\ as
\eqn\partph{\eqalign{
Z(\vec{J}, J_4) = \int_G & d^4A \ d^3E \
\d(\p_iA_i) \det(- D_i\p_i)    \cr
& \times \exp \int d^4x  \
[ iE_i(\dot{A}_i - D_iA_4) - (1/2)(E_i^2 + B_i^2)  - i g_0J_\m A_\m)],
}}
where
$B_1^a = \p_2 A_3^a - \p_3 A_2^a + g_0f^{abc}A_2^b A_3^c$ etc.
Integration on $A_4$ imposes color-Gauss's law, $D_i E_i = g_0 J_4$, in the
form of the constraint $\d(D_iE_i - g_0J_4)$,
\eqn\pgauss{\eqalign{
Z(\vec{J}, J_4) = \int_G & d^3A \ d^3E \
\d(\p_iA_i) \det(- D_i\p_i) \ \d(D_iE_i - g_0J_4)    \cr
& \times \exp \int d^4x  \
[ iE_i\dot{A}_i^{\rm tr} - (1/2)(E_i^2 + B_i^2) - i g_0 J_i A_i^{\rm
tr})]. }}
The constraint expressed by $\d(\p_iA_i)$
has allowed us to replace $A_i$ by its transverse part $A_i^{\rm tr}$
everywhere. We separate the transverse and longitudinal parts of $E_i =
E_i^{\rm tr} - \p_i \f$, and we have $d^3E = N dE^{\rm tr} d\f$.
The Faddeev-Popov determinant is absorbed by
\eqn\sgauss{\eqalign{
\det(- D_i\p_i) \ \d(D_iE_i - g_0 J_4) & =
\det(M) \ \d(M\f - \r_{\rm coul} - g_0 J_4)    \cr
   & = \d[\f - M^{-1}( \r_{\rm coul} + g_0 J_4)],
}}
where the symbols are defined as in Eq.\ \gslawa.  We now integrate over
$\f$, and in a similar way we integrate out the longitudinal part of $A_i$, to
obtain \pcoul.

\appendix B{One-loop expansion}

	In this Appendix we find the one-loop expansion of the quantities $V_0$
and $P_0$ defined in Eqs.\ \inst\ and \nst, and which appear in
$D_{0,44} = V_0 + P_0$.

	The Faddeev-Popov operator
is written $M(A^{\rm tr}) = M_0 + M_1(A^{\rm tr})$, where
$M_0 \equiv - \p_i^2$ is the negative of the Laplacian, and
$(M_1)^{ac} \equiv - g_0 f^{abc}A_i^{b,{\rm tr}} \p_i$.  The color-Coulomb
potential energy functional ${\cal V}(A^{\rm tr})$, defined in Eq.\ \pot,
reads
\eqn\rpot{\eqalign{
{\cal V}(\vx, \vec{y}; A^{\rm tr}) =
    [(M_0 + M_1)^{-1}M_0 (M_0 + M_1)^{-1}]_{\vec{x},\vec{y}},
}}
and has the expansion
\eqn\epot{\eqalign{
{\cal V}(\vx, \vec{y}; A^{\rm tr}) =
	[M_0^{-1} - 2 M_0^{-1} M_1 M_0^{-1}
+ 3 M_0^{-1} M_1 M_0^{-1} M_1 M_0^{-1} + \ldots]_{\vec{x},\vec{y}} \ ,
}}
where $M_{0}^{-1}|_{\vec{x},\vec{y}}
= (2\pi)^{-3}\int d^3k \ (\vk^2)^{-1} \exp[i\vk \cdot(\vx - \vec{y})]
= (4 \pi |\vx - \vec{y}|)^{-1}$. From Eq.\ \inst\
for $V_0$ we obtain to one-loop order
\eqn\expi{\eqalign{
V_0(x-y) = g_0^2 [ M_0^{-1} + 3 M_0^{-1}
\langle \ M_1 M_0^{-1} M_1 \ \rangle_0 \ M_0^{-1}]_{\vec{x},\vec{y}}
\ \d(x_4 - y_4),
}}
where we have used $\langle M_1(A^{\rm tr}) \rangle = 0$, which holds
because $M_1(A^{\rm tr})$ is linear in $A^{\rm tr}$.  The average
designated by $\langle \ldots \rangle_0$ refers to the free-field average, with
free-field propagators given in Eq.\ \prop.  This gives
\eqn\expi{\eqalign{
V_0 = V_{0,0} + V_{0,1}
}}
where the zero-loop piece is given explicitly by
\eqn\expiz{\eqalign{
V_{0,0}(x-y) \d^{ae} = g_0^2  M_0^{-1}(\vx - \vec{y}) \ \d(x_4 - y_4)
\ \d^{ae}
}}
and the one-loop piece by
\eqn\expio{\eqalign{
V_{0,1}(x-y) \d^{ae}= 3 g_0^4 \int d^3x' d^3y' \ &
f^{abc} f^{cde}
\langle A_i^{{\rm tr},b}(\vx', x_4)
A_j^{{\rm tr},d}(\vec{y}', x_4) \rangle_0 \cr
& M_0^{-1}(\vx - \vx')  \ \p_i M_0^{-1} (\vx' - \vec{y}')
\ \p_j M_0^{-1}(\vec{y}' - \vec{y})
\ \d(x_4 - y_4).
}}
These terms are illustrated in Fig.\ 1a.  In momentum space we have
$V_{0,0} = g_0^2/\vk^2$, and
\eqn\momo{\eqalign{
V_{0,1}(|\vk|) = { {3g_0^4N} \over {(\vk^2)^2}}
(2\pi)^{-4}\int d^4p \
{ {k_i(\d_{ij}-\hat{p}_i\hat{p}_j)k_j} \over
{(\vec{p}^{\, 2} + p_4^2)(\vec{p} - \vk)^2} }.
}}
The result of this integral is given in \peo\ and \cmp.

	Similarly, for $P_0$ given in Eq.\ \nst, we have to one-loop order
\eqn\expr{\eqalign{
P_0(x-y) \d^{ad} = - g_0^2 \langle \  (M_0^{-1}\r_{\rm coul}^a)(x)
\  (M_0^{-1}\r_{\rm coul}^d)(y) \ \rangle_0.
}}
where $\r_{\rm coul}^a = - g_0 f^{abc}A_i^{{\rm tr},b} E_i^{{\rm tr},c}$.
This gives
\eqn\expra{\eqalign{
P_0(x-y) \d^{ad}
= - \ & g_0^2  \int d^3x' d^3y' \ M_0^{-1}(\vx - \vx')  \cr
& \times
\langle \r_{\rm coul}^a(\vx', x_4) \r_{\rm coul}^d(\vec{y}', y_4) \rangle_0
 \ M_0^{-1}(\vec{y}' - \vec{y}),
}}
where
\eqn\expra{\eqalign{
\langle \r_{\rm coul}^a(x) \r_{\rm coul}^d(y) \rangle_0
= g_0^2 f^{abc} f^{deg}
[ &\langle A_i^{{\rm tr},b}(x) A_j^{{\rm tr},e}(y) \rangle_0
\langle E_i^{{\rm tr},c}(x) E_j^{{\rm tr},g}(y) \rangle_0   \cr
+ &\langle A_i^{{\rm tr},b}(x) E_j^{{\rm tr},g}(y) \rangle_0
\langle E_i^{{\rm tr},c}(x) A_j^{{\rm tr},e}(y) \rangle_0].
}}
This term is illustrated in Fig.\ 1b.  In momentum space it is given by
\eqn\momp{\eqalign{
P_{0,1}(k) = { {-g_0^4N} \over {(\vk^2)^2}}
(2\pi)^{-4}\int d^4p \
{ { P_{ij}(\vec{p}) } \over
{(\vec{p}^{\, 2} + p_4^2)} }
\ { { P_{ij}(\vec{p} - \vk) } \over
{[(\vec{p} - \vk)^2} + (p_4 - k_4)^2] }
[\vec{p}^{\, 2} - p_4(p_4 - k_4)],
}}
where $P_{ij}(\vec{p}) = \d_{ij} - \hat{p}_i \hat{p}_j$ is the
3-dimensionally transverse projector.  The contraction in the numerator
gives 2 terms,
\eqn\cntr{\eqalign{
 & P_{ij}(\vec{p})P_{ij}(\vec{p} - \vk) = J_1 + J_2  \cr
& J_1 = 2  \cr
& J_2 = - \ {{ \vec{p}^{\, 2}\vk^2 - (\vec{p} \cdot \vk)^2 }
\over {\vec{p}^{\, 2}(\vec{p} - \vk)^2 } } .
}}
Each term results in a Feynman integral $I_1$ and $I_2$.  The integral
$I_2$ looks more complicated.  However it is only logarithmically
divergent by power counting, and when the integration is performed,
the coefficient of the divergent part of $I_2$
vanishes, so $I_2$ is finite.  As a result $I_2(|\vk|, k_4)$ vanishes in
the limit $k_4 \to \infty$, and does not contribute to
$P_0^{\rm as}(k)$.  The result of the $I_1$ integration is given in
\peo\ and \olr.

	The integrals \momo\ and \momp\ are evaluated by dimensional
regularization, with $p_4 \to p_d$, and $\vec{p} =  (p_i)$ for $i =
1,\ldots, (d-1)$.

\vskip 2cm

\footatend\vfill\supereject\immediate\closeout\rfile\writestoppt
\baselineskip=14pt\centerline{{\bf References}}\bigskip{\frenchspacing%
\parindent=20pt\escapechar=` \input refs.tmp\vfill\eject}\nonfrenchspacing

%%%%%%%%%%%%%%%%%%%%%%%%%%%%%% FIGURES %%%%%%%%%%%%%%%%%%%%%%%%%

\vfill\eject\immediate\closeout\ffile{\parindent40pt
\baselineskip14pt\centerline{{\bf Figure Captions}}\nobreak\medskip
\escapechar=` \input figs.tmp\vfill\eject}

\topinsert
\vbox{
\vskip4.0 true cm
%\hskip-2.0 true cm
%
\centerline{
\epsfxsize=10.0 true cm
\epsfbox{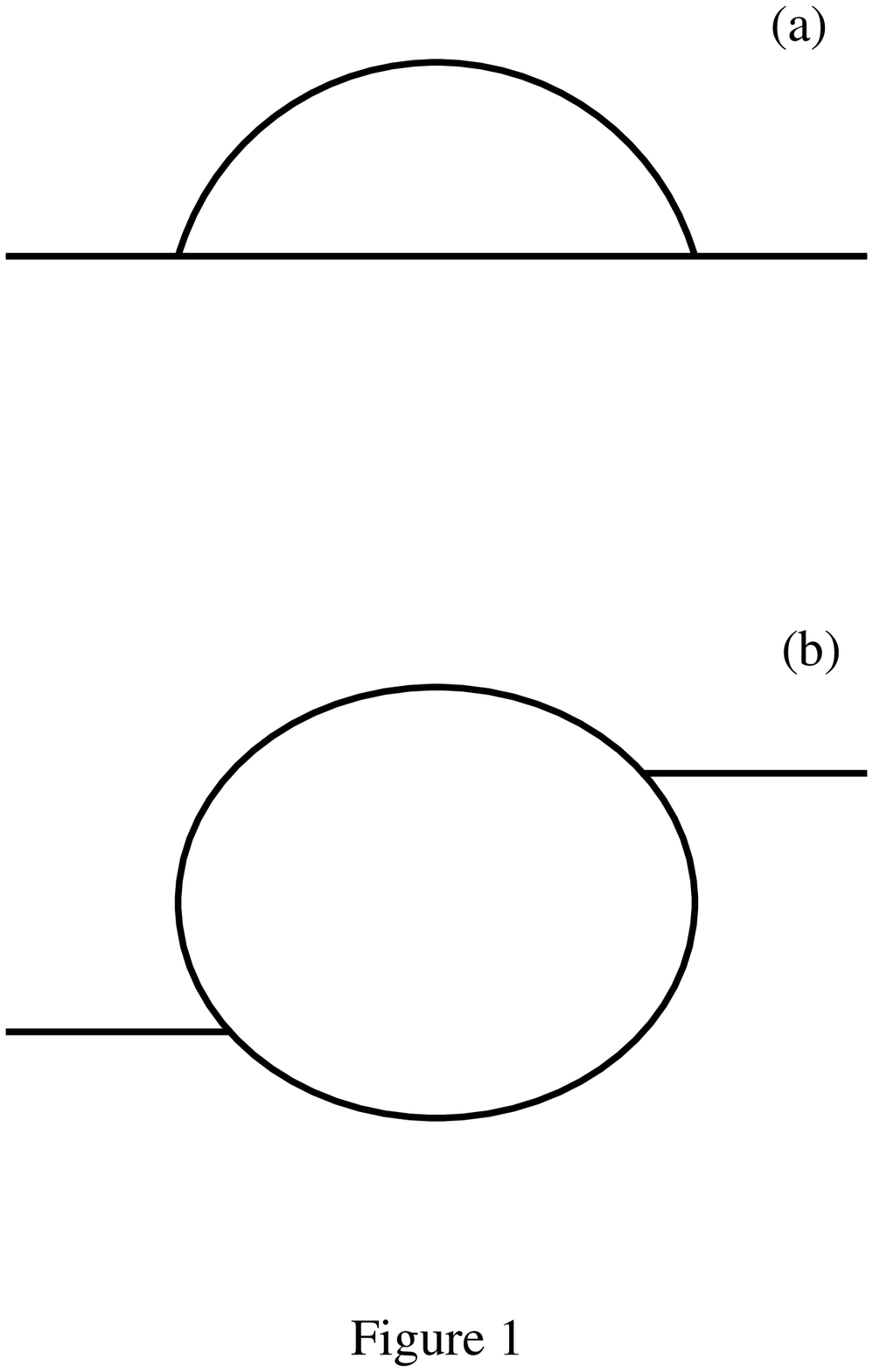}
}
% \vskip2.0 true cm
}
\endinsert

\bye